\documentclass[usenatbib]{mn2e}
\usepackage{graphicx}
\usepackage{epsfig}
\usepackage{subfigure}
\usepackage{cite}
\newcommand{\msun}{\>{\rm M_{\odot}}}

\newcommand{\beq}{\begin{equation}}
\newcommand{\eeq}{\end{equation}}

\newcommand{\Obaryon}{{\Omega_{\rm B,0}}}

\newcommand{\msunh}{\>h^{-1}\rm M_\odot}

\newcommand{\mvir}{m_{\rm vir}}
\newcommand{\rvir}{r_{\rm vir}}
\newcommand{\vvir}{v_{\rm vir}}

\newdimen\hssize
\newdimen\hdsize 
\def\vvir{V_{\rm vir}}

\usepackage{color}

\begin{document}

\title[Bayesian inference on star formation feedback] 
{Bayesian inferences of galaxy formation from the $K$-band
  luminosity and HI mass functions of galaxies: constraining star
  formation and feedback} \author[] {Yu Lu$^{1}$\thanks{E-mail:
    luyu@stanford.edu}, H.J. Mo$^{2}$, Zhankui Lu$^{2}$, Neal
  Katz$^{2}$, Martin D. Weinberg$^{2}$
  \\
  $^1$ Kavli Institute for Particle Astrophysics and Cosmology,
  Physics Department, \\ and SLAC National Accelerator Laboratory,
  Stanford University, Stanford, CA 94305, USA
  \\
  $^2$ Department of Astronomy, University of Massachusetts, Amherst
  MA 01003-9305, USA}


\date{}

\maketitle

\label{firstpage}


\begin{abstract}
  We infer mechanisms of galaxy formation for a broad family of
  semi-analytic models (SAMs) constrained by the $K$-band luminosity function
  and HI mass function of local galaxies using tools of Bayesian
  analysis. Even with a broad search in parameter space the whole model family
  fails to match to constraining data.  In the best fitting models,
  the star formation and feedback parameters in low-mass
  haloes are tightly constrained by the two data sets, and the
  analysis reveals several generic failures of models that
  similarly apply to other existing SAMs.
  First, based on the assumption that baryon accretion follows
  the dark matter accretion, large mass-loading factors are required
  for haloes with circular velocities lower than $200\,{\rm km/s}$,
  and most of the wind mass must be expelled from the
  haloes. Second, assuming that the feedback is powered by Type-II
  supernovae with a Chabrier IMF, the outflow requires more than 25\%
  of the available SN kinetic energy.  Finally, the posterior
  predictive distributions for the star formation history are dramatically
  inconsistent with observations for masses similar to or smaller than
  the Milky-Way mass.  The inferences suggest that the current model
  family is still missing some key physical processes that regulate
  the gas accretion and star formation in galaxies with masses below
  that of the Milky Way.
\end{abstract}


\section{Introduction}
\label{sec:intro}

The formation and evolution of galaxies is regulated by mass
accretion, star formation, and associated physical processes, such as
feedback from supernova (SN) explosions.  Unravelling this physical
complexity is one of the great challenges in modern astrophysics
\citep[see][for an overview]{Mo2010}. An important hurdle for any
theoretical model of galaxy formation is a natural explanation for the differing
shapes of the dark matter halo mass function and the galaxy luminosity
and stellar mass function.  The mass function of dark matter haloes,
$n(M)$, scales with halo mass as $n(M)\propto M^{-2}$ at the low-mass
end, in contrast to the observed luminosity function of galaxies at
low-$z$, $\Phi (L)$, which has a shallow faint-end slope: $\Phi(L)
\propto L^{-1}$. This suggests that star formation in low-mass haloes
must be inefficient, but the physical cause remains unexplained.
Similarly, the dependence of the baryonic mass fraction on halo mass
requires explanation.  One might expect that each halo has a baryonic
mass fraction that is close to the universal value of 17\%. In
reality, the total inventory of baryons in Milky Way-sized haloes
is only 5--10\% of the halo mass, and the baryon
mass fraction decreases rapidly with decreasing halo mass at the
low-mass end \citep[e.g.][]{Yang2003,Papastergis2012}.  Therefore,
some processes must keep a large fraction of baryons outside of
galaxies.

The process most often considered to suppress star formation in
low-mass haloes is SN feedback; the total amount of energy released by
supernovae can be larger than the binding energy of the gas in
low-mass haloes \citep{Dekel1986,White1992}.  Therefore, it is
energetically possible to expel large amounts of baryonic matter from
low-mass haloes and, thereby, reduce the efficiency of star formation.
Feedback may do this in multiple ways.  For example, the feedback
energy may heat a fraction of the cold ISM in the disk, either causing
it to mix with hot halo gas or ejecting it directly without
mixing \citep{Somerville1999, Benson2003}.  The feedback energy may
also interact with the halo gas, heat it, and then eject it from the
halo \citep{Croton2006}.  Alternatively, galactic winds may be driven
by radiation pressure associated with massive stars and supernova
explosions \citep[e.g.][]{Murray2005}. The total amount of energy in
radiation is two orders of magnitude larger than the available kinetic energy in
supernova explosions, and so the energy budget can be increased
significantly. Some simulations of isolated galaxies demonstrate that
radiation pressure-driven (also called momentum-driven) winds are
effective at ejecting cold gas from galaxies
\citep[e.g.][]{Hopkins2011b, Hopkins2013}, but other simulations
question their efficiency \citep[e.g.][]{Krumholz2012b}.  Furthermore,
cosmological simulations that employ simplified versions of the
momentum-driven model have not successfully reproduced the observed
galaxy mass function \citep[e.g.][]{Oppenheimer2006,Puchwein2012}, although
in the latest simulations the discrepancies are predominantly at high masses
where supernova winds may not be the dominant feedback mechanism 
\citep{Dave2013}. In
summary: 1) none of the feedback processes are understood in
detail; and 2) the effect of feedback on the star-formation efficiency is
poorly determined; the efficiency parameter adopted in the models 
ranges from a few percent to about 100\% \citep{Bower2012, Mutch2013},
with a large range of uncertainty.

Observationally, significant outflows are observed only for
starburst galaxies, and there is no clear evidence for massive winds
from normal spiral galaxies \citep[e.g.][]{Heckman2003}. The mass
loading and velocity of galactic winds are poorly constrained
\citep{Martin2005, Weiner2009, Chen2010}.  \citet{Martin2006}
estimated that mass outflow rates in the cold component of the wind is
on the order of 10\% of the star formation rate (SFR) in the
starburst, while other observations suggest that the rate may be as
high as or even higher than the SFR of the galaxy \citep{Martin2002}.

The feedback processes described above assume that baryons accrete
into a halo at the cosmic baryon fraction and that
a large fraction of the cooled baryons are subsequently ejected by feedback.  
Most semi-analytic models (SAMs) have focused on gas
ejection through winds and we will follow suit here.  Owing to the
uncertainty in the physical details, we use the SAM developed by
\citet{Lu2011a} that both incorporates sufficiently general
parametrisations of galaxy-formation mechanisms so that its results will
apply to a large class of SAMs and uses advanced Markov-Chain Monte
Carlo (MCMC) techniques to effectively explore the high dimensional
parameter space.  Similar approaches have been adopted by
\citet{Kampakoglou2008}, \citet{Henriques2009}, \citet{Henriques2013}, and \citet{Mutch2013}.  In
addition, Gaussian process model emulators have also been adopted to
explore the parameter space of SAMs by \citet{Bower2010} and \citet{Gomez2012}.
More generally, since the Bayesian inference approach
places the model on a probabilistic footing, comparisons between models,
probabilities within models, and the consistency of the models given the
data (i.e. goodness of fit) can all be assessed quantitatively.

In \citet{Lu2012}, we applied the Bayesian SAM to make model
inferences conditional on the observed $K$-band luminosity function of
local galaxies. We found that star formation and supernova (SN)
feedback are degenerate when one only uses the luminosity function to
constrain the model.  The model implied two modes for low-mass haloes: one
where the star formation efficiency is increasingly lower for lower halo masses
and one where the SN feedback is increasingly effective in reheating the cold
gas.  The dominating low star-formation mode hides a large
fraction of baryonic mass in the cold gas phase, thereby vastly over
predicting the HI mass function of galaxies. In this paper, we use
both the $K$-band luminosity function and the HI mass function of galaxies
to constrain a model family based on \citet{Lu2012}, with additional
feedback processes for ejecting baryons out of haloes,
to see if there are interesting regions in the extended 
parameter space that can explain both data sets simultaneously. 
We choose the $K$-band luminosity function because
it is less affected by dust extinction than are optical bands. The cold
gas component, although a relatively small fraction of the total
baryon gas \citep[e.g.][]{Zwaan2005, Giovanelli2005, Keres2003,
  Bell2003a}, depends crucially on gas cooling and accretion, star
formation, and feedback \citep{Wang2012, Lu2012} and thereby
complements the information from the stellar component.  We first use
our SAM with 17 free parameters to obtain the posterior distribution
from the constraining data. Our philosophy in this paper is to take a broad
prior range for the parameters, a range that is sometimes broader than
a conventional interpretation of observations and theory would allow, but one
that is broad enough to contain values assumed in published SAMs.
We find that using these very broad ranges for the priors
of the model parameters, our posterior distribution recovers many parameter
combinations of previously published SAMs, but the model is still unable to
match both of the constraining data sets simultaneously. If we restrict
our prior ranges to those suggested by other observations (and theory) then
the mismatch with the constraining observation becomes much worse,
as we discuss later.  We then analyse the posterior distribution
of the key parameters that govern star formation and feedback in an attempt
to gain insight into the underlying physical processes and to investigate
the shortcomings of the model.  Finally, we use the
constrained posterior to make predictions and compare them with
existing data to further test the models.

The paper is organised as follows.  \S\ref{sec:model} describes
the extended model based on \citet{Lu2011a, Lu2012}.  We present the
data, its error model, and the definition of the likelihood function
for the Bayesian inference in \S\ref{sec:data}. \S\ref{sec:inference}
presents the results of our model inference using
the constrained posterior distribution of model parameters
(\S\ref{sec:constraint}) and model predictions
(\S\ref{sec:predictions}).  Finally, we discuss our results in
\S\ref{sec:discussion}.  Throughout the paper, we assume a
$\Lambda$CDM cosmology with $\Omega_{\rm M,0}=0.26$, $\Obaryon=0.044$,
$h=0.71$, $n=0.96$, and $\sigma_8=0.79$, which are consistent with the
WMAP5 data \citep{Dunkley2009, Komatsu2009}.

\section{The semi-analytic model}
\label{sec:model}

We adopt the SAM developed by Lu et al. (2011; see also Lu et
al. 2012), which includes star formation, supernova (SN) feedback,
galaxy mergers and AGN feedback.  The processes are parametrised with
standard analytic functions so that the model encompasses a large range of
previously used prescriptions.  We only describe the parametrisations
relevant to SN feedback and star formation in this section; see
\citet{Lu2011a} for a comprehensive description of the other details.

Our model assumes that quiescent star formation occurs in a
high-density, exponentially-distributed cold-gas disk.  The
characteristic radius for the exponential disk, $r_{\rm disc}$, is
related to the virial radius and the spin parameter of the host halo
as follows: $r_{\rm disc}= \lambda\rvir/\sqrt{2}$, where $\lambda$ is
the spin parameter of the halo \citep{Mo1998}.  We choose a single
value $\lambda=0.035$ for all haloes; this value is roughly the
average value measured in cosmological $N$-body simulations
\citep[e.g.][]{Maccio2007}. The fiducial size of the disc is then
$r_{\rm disc,0}=0.035\rvir/\sqrt{2}$.  Observationally, gas forms
stars when its surface density exceeds a threshold, $\Sigma_{\rm SF}$;
the observed threshold surface density is $\sim 10\msun\, {\rm pc}^{-2}$
\citep[e.g.][]{Kennicutt1998, Kennicutt2007, Bigiel2008}.  For an
exponential disk, the cold gas mass with a surface density higher than the
threshold surface density $\Sigma_{\rm SF}$ is
\begin{equation}\label{equ:sf_msf_lu}
m_{\rm sf}=m_{\rm cold}\left[1-\left(1+\ln {m_{\rm cold} \over 2\pi f_{\rm SF}
       r_{\rm disc,0}^2}\right) {2\pi f_{\rm SF} r_{\rm disc,0}^2 \over m_{\rm cold}}\right]\,,
\end{equation}
where $m_{\rm cold}$ is the total cold gas mass in the galaxy, and $f_{\rm
  SF}$ is a model parameter defined as
\begin{equation}
f_{\rm SF}=\left({\lambda \over 0.035}\right)^2 \Sigma_{\rm SF}.
\end{equation}
The model parameter $f_{\rm SF}$ describes both the size of the disk
and the threshold surface density of the cold gas.  Thus, the
parameters $\lambda$ and $\Sigma_{\rm SF}$ are intrinsically
degenerate through the combined parameter $f_{\rm SF}$.  If $f_{\rm
  SF}$ is significantly lower than $10\msun {\rm pc}^{-2}$, then
either the cold gas disc is more compact than the fiducial size, or
the threshold surface density is required to be lower than $10 \msun
{\rm pc}^{-2}$. We assign a generous prior for the parameter $f_{\rm SF}$ 
to accommodate the broad uncertainties in the implementation 
of star formation in the model (see \ref{tab:param}), 
and we test the sensitivity of our results on the choice of 
the prior, which will be discussed in \S\ref{sec:constraint}. 

We assume that the star formation rate, $\phi$, is proportional to the mass of
star forming gas and inversely proportional to the dynamical time
scale of the disk, $\tau_{\rm disc}={r_{\rm disc}/\vvir}$:
\begin{equation}\label{equ:sf_lu}
\phi=\epsilon_{\rm sf}{m_{\rm sf} \over \tau_{\rm disc}},
\end{equation}
where $\epsilon_{\rm sf}$ is an overall star formation efficiency factor.  To
further generalise the dependence of the star formation rate on halo
properties, the model further assumes that $\epsilon_{\rm sf}$ has a
broken power-law dependence on the circular velocity of the host halo:
\begin{equation}\label{equ:sf_ef_lu}
  \epsilon_{\rm sf} = \left\{ \begin{array}{ll}
      \alpha_{\rm SF} & \mbox{$\vvir \geq V_{\rm SF}$}; \\
      \alpha_{\rm SF}\left({\vvir \over V_{\rm SF}}\right)^{\beta_{\rm SF}}
      & \mbox{$\vvir < V_{\rm SF}$}, \end{array} \right. 
\end{equation}
where $\alpha_{\rm SF}$, $\beta_{\rm SF}$ and $V_{\rm SF}$
characterise the star formation efficiency. For haloes with a circular velocity
higher than the critical velocity scale, $V_{\rm SF}$, the star formation 
efficiency is a constant, $\alpha_{\rm SF}$; for haloes with a circular velocity
lower than $V_{\rm SF}$, the star formation efficiency follows a power-law
function of halo circular velocity defined by $\alpha_{\rm SF}$ and 
$\beta_{\rm SF}$.  The pivot point $V_{\rm SF}$ in the parametrisation is
needed to avoid unrealistically high star formation efficiencies in high
$\vvir$ haloes when $\beta_{\rm SF}$ is positive.
However, we note that it introduces an intrinsic degeneracy into the 
parametrisation.  For example, when $\beta_{\rm SF}=0$, $V_{\rm SF}$ does not 
have any impact on the model.  This effects our study, as we 
discuss in \S\ref{sec:constraint}.  We note that allowing
$\beta_{\rm SF}$ to have nonzero values conflicts with some observational
results,  which find that the star-formation rate is largely determined by the
available supply of cold dense gas and not on the mass (or
circular velocity) of the host halo \citep{Schmidt1959, Kennicutt1989}.

Star formation in our SAM can both consume cold gas by forming stars
and affect the cold gas supply for subsequent star formation 
through feedback.  In general, star formation feedback 
can influence the cold gas in three different ways:
first, feedback can heat up the cold gas in the disk so that it joins
the hot halo; second,  feedback can directly eject cold gas out of the 
gravitational potential of the host halo; and third, feedback can expel some
of the hot halo gas out the host halo. 
Our model attempts to capture all these processes. 
In our previous studies \citep{Lu2011a}, we did not explore
the dependence on cold-gas ejection 
owing to its strong degeneracy with other mechanisms when the luminosity
function or stellar mass function alone is used as a constraint.  The
addition of the HI mass function breaks this degeneracy, and we will
see that cold-gas ejection is required by the HI data.  Our previous
papers enforced conservation of the mechanical energy produced by SN
explosions for the chosen stellar initial mass function (IMF) when
modelling the mass loading in galaxy winds.  In this paper, we allow
the possibility of other energy sources, e.g. radiatively-driven winds.

We parametrise the mass-loss rate of the cold gas to be proportional to the 
star formation rate, $\phi$, as
\begin{equation}
\dot{m}_{\rm out}=\alpha_{\rm ld} \phi\,,
\end{equation}
where the coefficient $\alpha_{\rm ld}$ is the loading factor of the
outflow.  Because the quantity of energy or momentum produced during
stellar evolution is finite, the mass-loading depends on the initial
wind velocity and the physics of the coupling with the multiphase ISM.
Two types of mass-loading models have been extensively explored in the literature: 
1) an energy-driven wind that couples a fixed fraction of the feedback
energy with the wind, which motivates $\alpha_{\rm ld}\propto v_{\rm
  w}^{-2}$ \citep[e.g.][]{Okamoto2010}; and 2) a momentum-driven wind
that couples a fixed fraction of momentum (e.g. in a radiation field)
with the wind, which motivates $\alpha_{\rm ld}\propto v_{\rm w}^{-1}$
\citep[e.g.][]{Oppenheimer2006, Oppenheimer2010, Hopkins2013}.
Attempts to observe $v_{\rm w}$ and infer its dependence on galaxy
properties \citep[e.g.][]{Martin1999} have been inconclusive.  Some
theoretical models assume a constant wind velocity
\citep[e.g.][]{Croton2006}, while others assume a halo mass dependent
velocity \citep[e.g.][]{Puchwein2012}.  To parametrise the various
possibilities, we assume the loading factor to be a power-law function
of halo circular velocity, $\vvir$,
\begin{equation}
\alpha_{\rm ld}=\alpha_{\rm LD} 
\left({\vvir \over V_0} \right)^{-\beta_{\rm LD}},
\end{equation}
where the power index $\beta_{\rm LD}$ and the normalisation
$\alpha_{\rm LD}$ are model parameters, and $V_0$ is an arbitrary
velocity scale that we fix at $220\,{\rm km/s}$.  

In this model, a given value of $\beta_{\rm LD}$ does not determine a
particular wind launching mechanism (energy driven or momentum
driven).  For example, $\beta_{\rm LD}=0$ implies independence of
$\vvir$, but this does not affect the $\alpha_{\rm ld}$-$v_{\rm w}$ 
relation predicted by a particular wind driving mechanism, because 
the $v_{\rm w}-\vvir$ relation is undetermined.
Only in the special case when $v_{\rm w}\propto \vvir$ is the value
of $\beta_{\rm LD}$ related to the wind driving mechanism: $\beta_{\rm
  LD}=2$ for an energy-driven wind and $\beta_{\rm LD}=1$ for a
momentum-driven wind. Because of all the uncertainties, 
we assign a large prior range for $\beta_{\rm LD}$, allowing $\beta_{\rm LD}$ to
take any value between 0 and 8 in our inference. 

After the gas is removed from the galaxy, it can be trapped within the
potential well of the dark matter halo or ejected from the halo into
the intergalactic medium (IGM).  We assume that the fraction of removed gas
that is ejected from the halo is 
\begin{equation}
f_{\rm ej}=\left[1+\left({ \vvir 
\over V_{\rm EJ}} \right)^{\beta_{\rm EJ}}\right]^{-1},
\end{equation}
where $V_{\rm EJ}$ and $\beta_{\rm EJ}$ are free parameters.  Thus,
for a halo with a circular velocity lower than $V_{\rm EJ}$, most of
the outflow gas is ejected from the halo, while for haloes with
circular velocities much larger than $V_{\rm EJ}$, the ejected
fraction follows a power-law function of the halo circular velocity.
Our model also includes an outflow of hot halo gas. The strength of the effect
is controlled by the parameter $\alpha_{\rm SN}$, 
which describes the fraction of the total kinetic energy released 
by SN Type II that is needed to power all the feedback processes 
in the model. The excess energy over  the amount used 
for reheating and ejecting the cold gas is used to power an 
outflow of hot halo gas. 
The amount of halo gas mass expelled is written as
\begin{equation}\label{equ:wind_lu}
\Delta m_{\rm wind} = \epsilon_{\rm W}   
  \left\{ \alpha_{\rm SN}\frac{V_{\rm SN}^2}{V_{\rm esc}^2} -
  \alpha_{\rm ld} \left[\left(\frac{\vvir}{V_{\rm esc}}\right)^2+f_{\rm ej}\right]\right\}
   \phi  \Delta t~,
\end{equation}
where $V_{\rm esc}$ is the escape velocity of the halo. For a NFW halo
with a concentration $c$ \citep{Navarro1996}, $V_{\rm esc}=\vvir
\times \sqrt{ 2c \over \ln (1+c) -{c \over 1+c}}$ \citep{Puchwein2013}.

The model assumes that the ejected gas can
re-collapse into the halo at later times as hot halo gas. 
As in \citet{Springel2001} and \citet{DeLucia2007}, the rate of
re-infall of ejected gas is given by
\begin{equation}\label{equ:fb_ri}
\dot{m}_{\rm re-infall}=f_{\rm RI} \left( { M_{\rm ej} \over \tau_{\rm dyn}} \right).
\end{equation}
Here, $f_{\rm RI}$ is a free parameter, $M_{\rm ej}$ is the total
mass of gas in the `ejected' reservoir, including both ejected gas from the cold gas disk 
and expelled gas from the hot halo, and $\tau_{\rm dyn} =
\rvir/\vvir$ is the dynamical time of the halo.

To convert the stellar mass to stellar light, we apply the stellar
population synthesis (SPS) model of \citep[BC07, see][]{Bruzual2007},
which includes an improved treatment of thermally-pulsing AGB stars.
The SPS model is used to predict the $K$-band luminosity of a galaxy
from its star formation history.  To implement the SPS model, we
use a Chabrier IMF \citep{Chabrier2003}, and create a look-up
table for the luminosities on a grid with 220 values of age evenly
spaced from 0 to 15~Gyr, and with $Z=0.0001$, $0.0004$, $0.004$,
$0.008$, $0.02$ and $0.05$.

Finally, our SAM uses Monte-Carlo-generated halo merger trees tuned to
match the conditional mass functions found in $N$-body simulations
following \citet{Parkinson2008} with final masses
in the range from $10^{9}\msunh$ to $10^{15}\msunh$ \citep[see][for
additional details]{Lu2011a}.  Since haloes and their merger trees are
randomly sampled from the halo mass function, a model prediction 
based on a finite merger tree sample suffers from sampling variance.  
We, therefore, choose the number of halo merger trees in each mass 
bin such that the standard deviation in the predictions, namely the 
$K$-band luminosity function and the HI-mass function, are at least 2 times 
smaller than the error in the observational data in all the bins.  
Specifically, we use 1,000 merger trees for haloes with present masses 
in the range $10^{11}$ -$10^{12.5}\msunh$, 1,500 in $10^{12.5}$ - $10^{13.5}\msunh$, 
400 in $10^{9}$ - $10^{10}\msunh$, 400 in $10^{10}$ - $10^{11}\msunh$, and
100 in $10^{13.5}$ - $10^{15}\msunh$.  The logarithmic halo masses are evenly 
distributed in each range. Since massive haloes are rare
in the assumed cosmology, their contribution to scatter in the
stellar mass function is negligible. We choose a mass resolution for the merger
trees that varies with the final halo mass. For haloes with
final masses smaller than $10^{10}\msunh$, the mass resolution is
$10^{8.5}\msunh$; for haloes with final mass between $10^{10}\msunh$
and $10^{12}\msunh$, the mass resolution is $10^{9.2}\msunh$; for
haloes with final mass between $10^{12}\msunh$ and $10^{14}\msunh$,
the mass resolution is $10^{10.2}\msunh$; and for haloes with final
masses larger than $10^{14}\msunh$, it is $10^{11}\msunh$. All merger
trees are sampled at 100 redshifts equally spaced in $\log(1+z)$ from
$z=7$ to $z=0$. To make predictions for the total galaxy population we
weight each predicted galaxy by the halo mass function of
\citet{Sheth2001}.
The merger tree set we adopt allows us to run each model rapidly with 
sufficient accuracy so that we can explore the parameter space within 
a reasonable amount of time. After we obtained the posterior, we checked our
model predictions using a larger number of trees and higher time and mass
resolutions.  The deviations of the results from those obtained with the
smaller tree sample and lower resolutions described above were
found to be well within the 1-$\sigma$ range of the posterior
predictive distributions, demonstrating that the number of trees and
mass resolution that we adopt are adequate for our purposes.

\begin{table*}
\caption{Model parameters} 
\centering
\begin{tabular}{l c c c c}
\hline\hline
\# & Parameter & Meaning & Prior \\
\hline
1 & $\log M_{\rm CC} (\msun) $ & cooling cut-off halo mass& [1.5 , 4.5] \\
\hline
2 & $\log \alpha_{\rm SF}$ & star formation efficiency power-law amplitude & [-3, 0] \\
\hline
3 & $\beta_{\rm SF}$ & star formation efficiency power-law index &  [-1, 12] \\
\hline
4 & $\log V_{\rm SF} $ (km/s) & star formation law turn-over halo circular velocity & [1.5, 3.0] \\
\hline
5 & $\log f_{\rm SF} $ & star formation threshold gas surface density & [0, 1.2] \\
\hline
6 & $\log \alpha_{\rm SN}$ & SN feedback energy fraction & [-3, 1] \\
\hline
7 & $\log \alpha_{\rm LD}$ & SN feedback reheating power-law amplitude & [-3, 2] \\
\hline
8 & $\beta_{\rm LD}$ & SN feedback reheating power-law index & [0, 14] \\
\hline
9 & $\log V_{\rm EJ}$ & SN feedback ejection pivot halo circular velocity & [1.6,3.0] \\
\hline
10 & $\beta_{\rm EJ}$ & SN feedback ejection power-law index & [0,10] \\
\hline
11 & $\epsilon_{\rm EJ}$ & SN feedback ejection escaping velocity factor & [1, 5] \\
\hline
12 & $\log \epsilon_{\rm W}$ & fraction of surplus SN feedback energy used for powering wind & [-3, 0] \\
\hline
13 & $\log f_{\rm RI}$ & fraction of re-infall ejected hot gas & [-2, 0] \\
\hline
14 & $\log f_{\rm DF}$ & merging time-scale in dynamical friction time-scale & [0, 2] \\
\hline
15 & $\log \alpha_{\rm SB}$ & merger triggered star burst efficiency power-law amplitude& [-2, 0] \\
\hline
16 & $\beta_{\rm SB}$ & merger triggered star burst efficiency power-law index & [0, 2] \\
\hline
17 & $\arctan(\alpha_{\rm IN})$ & $K$-band luminosity function faint-end incompleteness & [0, 0.177] \\
\hline
\hline 
\end{tabular}
\label{tab:param}
\end{table*}


\section{Data and likelihood}
\label{sec:data}

Our SAM inference is conditional on both the $K$-band luminosity
function and HI mass function of galaxies to constrain models. We use
the $K$-band luminosity function from \citet{Bell2003}; the details of
the data and how it is used to constrain the model can be found in
\citet{Lu2012}. Different from the luminosity function, whose errors can be safely approximated 
by Poisson errors, the HI mass function contains errors that are correlated 
across different bins. Here, we give a detailed description for our use of the
HI mass function from \citet{Zwaan2005}, derived from the HIPASS HI
Bright Galaxy Catalogue \citep[hereafter HIPASS BGC,
see][]{Koribalski2004}, and show how we derive a full covariance matrix for 
the likelihood function utilised in our Bayesian analysis. 

\subsection{Uncertainties and covariance in the observed HI mass function}
\label{sec:mass_function_uncertainties}

A galaxy's HI mass is estimated from its 21-cm flux density and
distance as
\begin{equation}
  M_{\rm HI}=2.36 \times 10^5 d^2F_{\rm HI} \msun,
\end{equation}
where $d$ is the distance of the source in Mpc, and $F_{\rm HI}$ is
the integrated 21-cm flux density in units of Jansky per km/s.  The
value of $F_{\rm HI}$ is provided for each galaxy in the catalogue,
along with its uncertainty, $\sigma (F_{\rm HI})$.  We use the
Hubble distance,
\begin{equation}
  d=v_{\rm LG}/H_0, 
\end{equation}
where $H_0$ is the Hubble constant in units of ${\rm km\,s^{-1}\,Mpc^{-1}}$, 
$v_{\rm LG}$ is the recession
velocity of the source, $v_{\rm sys}$, corrected to the Local Group
rest frame:
\begin{equation}
  v_{\rm LG}=v_{\rm sys} + 300 \sin l \cos b,
\end{equation}
and $l$ and $b$ are the Galactic longitude and latitude of the source,
respectively. The observational error in $v_{\rm sys}$ is provided in
the source catalogue and is used in our analysis.
 
The detectability of a HI galaxy is affected by both its peak flux
density, $S_{\rm p}$, and its linewidth, $w_{20}$, defined to be the
wavelength difference between the two points where the flux density is
20\% of $S_{\rm p}$. Following \citet{Koribalski2004}, we estimate the
uncertainty in the peak flux density as
\begin{equation}
  \sigma(S_{\rm p})^2= {\rm rms}^2 + (0.05 S_{\rm p})^2,
\end{equation}
where the ${\rm rms}=13\,{\rm mJy}$. The uncertainty in $w_{20}$ 
is estimated as $\sigma(w_{\rm 20})\approx 3 \sigma(v_{\rm sys})$. 

Using these values of $v_{\rm LG}$, $F_{\rm HI}$, $S_{\rm p}$, and
$w_{20}$ and their standard deviations, we randomly generate 1,000
replicas containing 1,000 galaxies, each assuming independent Gaussian
distributions.  For each of the replicas, we adopt the same procedure
used by \citet{Zwaan2005} to obtain the HI mass function.
Specifically, we apply the 2-dimensional, step-wise maximum likelihood
method (2DSWML) proposed by \citet{Zwaan2003} to find the maximum
likelihood (ML) estimator of the mass function. These ensembles yield
the average mass function and the corresponding covariance matrix,
$\Sigma_{\rm obs}$.  This covariance matrix includes both
HI-measurement variance and the sampling variance owing to the finite
survey volume. We will assume that both types of variance are
independent and use this covariance matrix to specify the likelihood
function in \S\ref{sec:likelihood}.

\subsection{Corrections for molecular fraction and incompleteness}
\label{sec:mass_function_corrections}

Our SAM predicts a galaxy's total cold mass, $M_{\rm cold}$, which
includes atomic HI gas, molecular ${\rm H}_2$ gas, as well as heavier
elements. We make the following assumptions to predict the HI mass function. 
To start, we write
\begin{equation}\label{eq_Mcold}
  M_{\rm cold}={ M_{\rm HI} + M_{\rm H_2} \over \beta},
\end{equation}
where $\beta\approx 0.74$ is used to correct for the mass in helium
(He) and in the small fraction of heavier elements in a cosmic
gas. Given the molecular to atomic hydrogen mass ratio, $\eta=M_{\rm
  H_2}/M_{\rm HI}$, we have
\begin{equation}
M_{\rm HI}={\beta M_{\rm cold} \over 1+\eta}\,.
\end{equation}
The value of $\eta$ depends on galaxy properties.  Following
\citet{Obreschkow2009}, we assume that $\eta$ depends only on the total
mass of cold gas,
\begin{equation}
  \log \eta = q + k \log \left({M_{\rm cold} \over 10^9 h^{-2} \msun} \right) + \sigma,
\end{equation}
where $q=-0.51^{+0.03}_{-0.04}$, $k=-0.24^{+0.05}_{-0.05}$, and
$\sigma=0.30$.  The HI mass derived from the observed 21-cm flux is
only a fraction of the total HI mass owing to self-absorption.
Although the details of self-absorption depend on inclination, the
average correction expected for a sample of randomly oriented galaxies
is about 15\% \citep{Zwaan2003}.
This gives a relation between the predicted, observed HI mass,
$M^p_{\rm HI}$, and the real HI mass $M_{\rm HI}$:
\begin{equation}
  M^p_{\rm HI}=(1-f_{\rm HI})M_{\rm HI}=\beta{1-f_{\rm HI} 
    \over 1+\eta} M_{\rm cold}.
\end{equation}
We take the self-absorption correction factor, $f_{\rm HI}$, to be a
random number between 0 and 0.3.

The detection of HI galaxies is also limited by a minimal velocity
width $w_{\rm m}$, below which galaxy signals cannot be reliably
distinguished from radio-frequency interference. Thus, galaxies with
low inclinations and low HI mass, which thereby have low velocity
widths, are under sampled.  In the HIPASS BGC, this
effect is specified in terms of the velocity width at 50\% of the peak
flux, 
\begin{equation}
w_{50}=(w_0^2 \sin^2 i + w_{\rm t}^2)^{1/2} + w_{\rm i}\,,
\end{equation}
where $i$ is the inclination, $w_0$ is the intrinsic velocity spread owing
to rotation, $w_{\rm t}$ is the velocity width owing to turbulence, and
$w_{\rm i}$ is the broadening owing to the instrument.  Following
\citet{Zwaan2003}, we set $w_{\rm t}=6\ {\rm km \,s^{-1}}$ and $w_{\rm
  i}=2.3\ {\rm km \, s^{-1}}$.  Galaxies included in the HIPASS BGC
have $w_{50}>w_{\rm m}=26.4\ {\rm km \, s^{-1}}$, so the minimal reliably
detected inclination is
\begin{equation}
  i_{\rm m} = \arcsin \left[{(w_{\rm m}-w_{\rm i})^2 - w_{\rm t}^2 \over w_0^2} \right]^{1/2}.
\end{equation}
The fraction of galaxies potentially missed at a given $w_0$ is then
\begin{equation}\label{equ:zeta}
  \zeta=\int_0^{i_{\rm m}} \sin i {\rm d}i = 1-\cos i_{\rm m} =
  1-\left[ 1- {(w_{\rm m}-w_{\rm i})^2 - w_{\rm t}^2 \over w_0^2}
  \right]^{1/2}.
\end{equation}
\citet{Zwaan2003} used the HI Tully-Fisher relation from
\citet{Lang2003} to infer $w_0=0.35 M_{\rm HI}^{0.3}$, ignoring the
scatter.  This implies that $\zeta\approx
10\%$ for $M_{\rm HI}=2\times 10^7 \msun$ and $\zeta \approx 5\%$ for
$M_{\rm HI}=5\times 10^7 \msun$.  We include this effect in our model
predictions for the HI mass function as
\begin{equation}
\Phi^p(M^p_{\rm HI}) = \Phi(M^p_{\rm HI}) (1-\zeta),
\end{equation}
with $\zeta$ given by  equation (\ref{equ:zeta}).

\subsection{The likelihood function}
\label{sec:likelihood}

For the $K$-band luminosity function, we use the same likelihood
function as described in \citet{Lu2012}.  For the HI mass function, we
construct a likelihood function that independently represents the
HI-measurement error and the sample variance.  To start, we neglect
the covariance between measurement error and sampling variance. Such
covariance may be important for small sample sizes but is unimportant
for the current sample where the variance is measurement-dominated
(Papastergis, E. private communication).  The uncertainties in the
process leading to the HI mass estimate for each galaxy 
(\S\ref{sec:mass_function_uncertainties}--\S\ref{sec:mass_function_corrections})
induces correlations between the HI mass bins.  Ideally, the error
model describing this covariance would be used to predict the observed
data.  However, our galaxy formation model is not capable of
predicting all the necessary observables (e.g. $w_{20}$ and $S_{\rm
  p}$). Instead, we proceed by making two additional assumptions: 1)
the likelihood function is a multi-dimensional normal distribution;
and 2) the sampling is a pure Poisson process.  Then, the errors in
the simulated data presented in \S\ref{sec:mass_function_uncertainties} 
can be modelled as follows.  We
define the equivalent volume, $V_{\rm eq}$, implied by the HI measurement
process including all the selection effects described in 
\S\ref{sec:mass_function_corrections}.  This equivalent volume can be
estimated from the data as $V_{\rm eq}=N_{\rm obs}/\mathbf{\Phi}_{\rm
  obs}$. The Poisson variance in each modelled bin is then
$\mathbf{\Phi}_{\rm mod}/V_{\rm eq}$ and only affects the diagonal
terms of the covariance matrix.  We, therefore,
write the diagonal terms as
\begin{equation}
  \sigma_{\rm mod}^{2} = \sigma_{\rm mst}^{2} + 
  \left(\frac{\Phi_{\rm mod}}{V_{\rm eq}} \right)
  \label{eq:sigmod}
\end{equation}
where $\sigma_{\rm mst}^{2}$ is some unknown measurement variance.
Under the same assumptions, the diagonal elements of the simulated
covariance matrix from \S\ref{sec:mass_function_uncertainties}
are
\begin{equation}
  \sigma_{\rm obs}^{2} = \sigma_{\rm mst}^{2} + 
  \left(\frac{\Phi_{\rm obs}}{V_{\rm eq}} \right)
  \label{eq:sigmobs}
\end{equation}
We now use equations (\ref{eq:sigmod}) and (\ref{eq:sigmobs}) to eliminate
$\sigma_{\rm mst}^{2}$:
\begin{equation}
  \sigma_{\rm mod}^2=\sigma_{\rm obs}^2 + { \mathbf{\Phi}_{\rm
      mod} - \mathbf{\Phi}_{\rm obs} \over V_{\rm eq}}.
\end{equation}
This replaces the variance in the diagonal terms of the simulated
covariance matrix to obtain the predicted covariance
$\mathbf{\Sigma}_{\rm pred}$.  Thus, the likelihood function is written as 
\begin{eqnarray}
  && L(\mathbf{\Phi}_{\rm obs}| \mathbf{\Theta}) = \frac{L_0}{(2
    \pi)^{I/2} |{\rm det}(\mathbf{\Sigma}_{\rm pred})|^{1/2}} \times
  \nonumber \\  && \exp\left[-{1 \over 2} (\mathbf{\Phi}_{\rm obs} 
    - \mathbf{\Phi}_{\rm mod})^T \cdot
    \mathbf{\Sigma}^{-1}_{\rm pred} \cdot (\mathbf{\Phi}_{\rm obs} -
    \mathbf{\Phi}_{\rm mod})\right]
    \label{eq:Lcov}
\end{eqnarray}
where $\mathbf{\Theta}$ denotes the model parameter vector,
$\mathbf{\Phi}_{\rm obs}$ and $\mathbf{\Phi}_{\rm mod}$ are,
respectively, the vectors of the observed and predicted HI mass
functions over the HI mass bins, and $I$ is the rank of the covariance
matrix.  We assume that the $K$-band luminosity function and the HI mass
function are statistically independent and, therefore, the total likelihood is
simply the product of the two:
\begin{equation}
  L(D|\mathbf{\Theta})
  =L(D_{\rm K}|\mathbf{\Theta})\times L(D_{\rm HI}|\mathbf{\Theta})\,,
\end{equation}
where $D_{\rm K}$ and $D_{\rm HI}$ denote the observed data of the $K$-band
luminosity function and the HI mass function, respectively.

\section{Model inferences}
\label{sec:inference}

\subsection{Inference from the Posterior Distribution}
\label{sec:constraint}

We used a tempered version of the Differential Evolution algorithm
\citep{Braak2006} implemented in the Bayesian Inference Engine
\citep[BIE, see][]{Weinberg2013} to sample the posterior density
distribution with 256 chains.  All simulations run for more than 7000
iterations and the convergence is monitored with the Gelman-Rubin
${\hat R}$ test \citep{Gelman1992}.  We declare convergence when
${\hat R}\la 1.2$.  All simulations typically converge after 3000
iterations, and the converged states are used to sample the posterior
distribution. After removing 12 outlier chains, we obtain $\sim 10^6$ chain
states. As the auto-correlation length 
of the chains is typically $\sim 10$, there are about $10^5$ independent 
chain states to sample the posterior distribution. 

\begin{figure}
 \begin{minipage}[t]{.5\textwidth}
   \centering
   \includegraphics[width=1\textwidth]{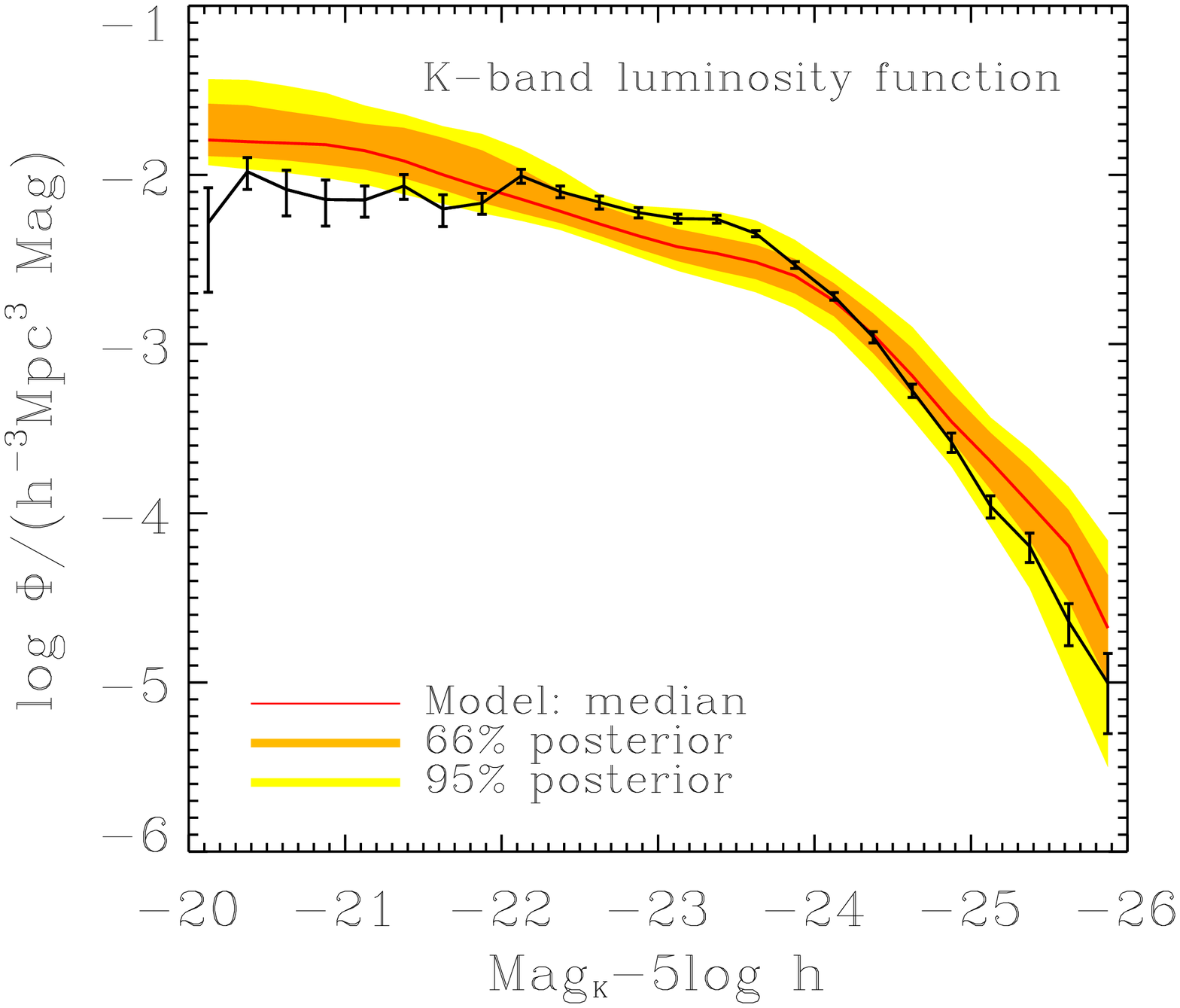}
 \end{minipage}
 \begin{minipage}[t]{0.5\textwidth}
   \centering
   \includegraphics[width=1\textwidth]{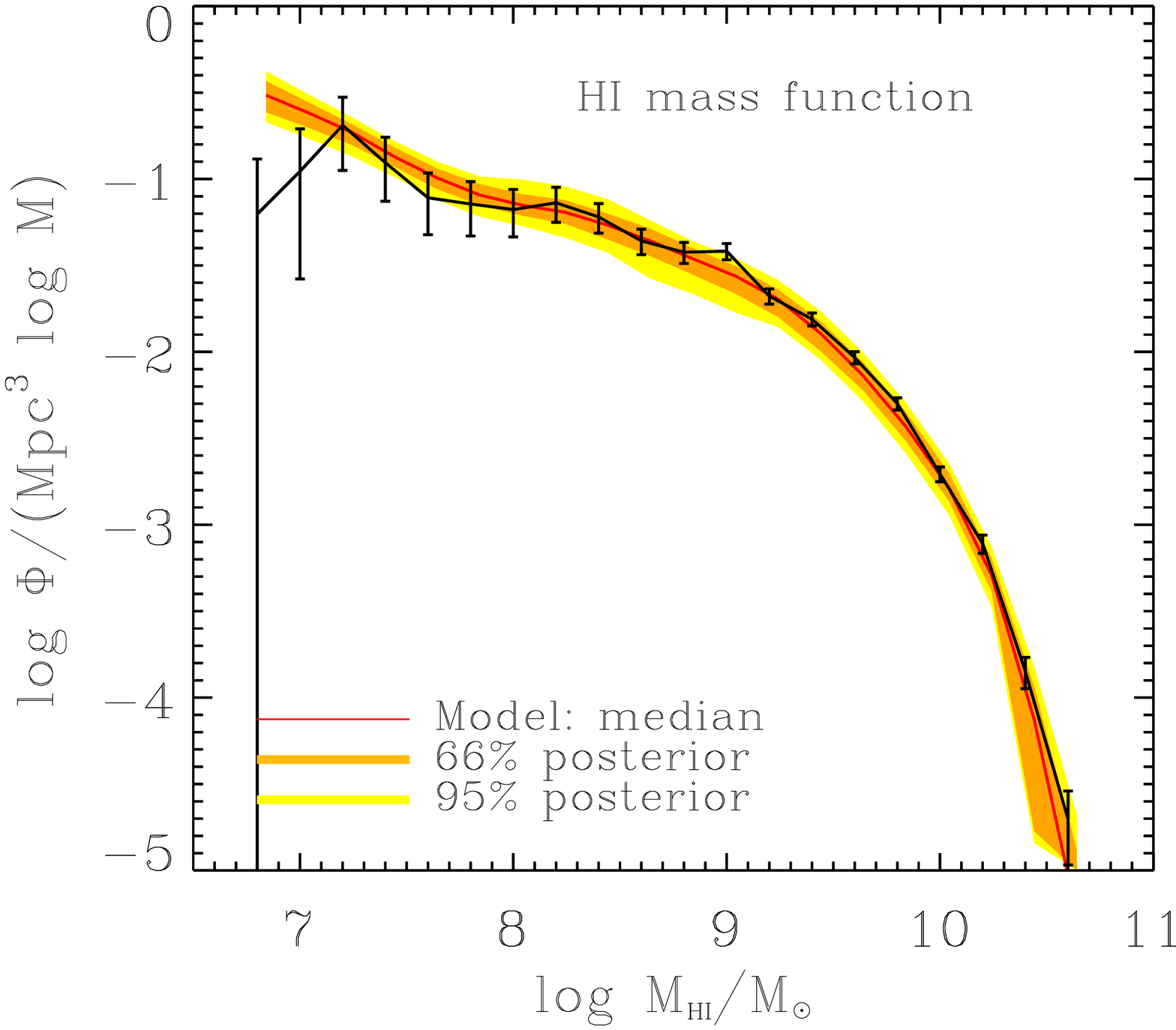}
 \end{minipage}
 \caption{The Bayesian posterior predictions for the $K$-band
   luminosity function and the HI mass function.  The
   black solid lines with error bars are the observational data.  The
   orange and yellow bands encompass the 66\% and 95\% credible
   ranges of the predictions, respectively, while the red lines are
   the median values.}
 \label{fig:constraints}
\end{figure}

Figure \ref{fig:constraints} shows the posterior predictions for the
$K$-band luminosity function and HI mass function of local galaxies.
We account for potential incompleteness in the faint end of the $K$-band
luminosity function by marginalising over the 
incompleteness parameter $\alpha_{\rm IN}$ \citep[see][for details]{Lu2012}.
The orange and yellow bands in each panel encompass 66\% and 95\% of
the posterior probability, respectively, and the red line is the
median.  To quantify the goodness of fit, we use \emph{posterior
  predictive check} \citep[PPC,][]{Rubin1984,Gelman1996,Gelman2003}.
PPC compares the value of one or more discrepancy statistics for
the observed data to a reference distribution based on the posterior
(which is conditioned on the observed data).  The reference
distribution is obtained by generating predicted data from the
posterior distribution. If a model fits the data well, the observed data should be
likely under the posterior predictive distribution, and
the discrepancy statistics of the observation should be likely 
under the distribution of the discrepancy statistics of 
the posterior models.  On the other hand,
large discrepancies between the observed data and the posterior
predictive distribution indicate that the model performs poorly.  
Following our previous paper \citep{Lu2012}, we adopt the 
Bayesian $p$-value to quantify the goodness of fit, 
with $p$ defined as
\begin{equation}
p = {1 \over N} \sum_{i=1}^N I_{\chi_{i, {\rm mod}}^2
  \geq \chi_{\rm obs}^2}\,,
\end{equation}
where $\chi_{i, {\rm mod}}^2$ and $\chi_{\rm obs}^2$ are the discrepancy statistics for the model and the data. 
$I_{q}$ is the indication function for the condition $q$, 
with $I_q=1$ if $q$ is true and $I_q=0$ otherwise.
In other words, the fraction of posterior model samples with 
$\chi_{i, {\rm mod}}^2 \geq \chi_{\rm obs}^2$ is an estimate of $p$. 
A small value of the $p$-value reflects the implausibility of the
data under the model (and, hence, the lack of fit of the model to the
data) and, therefore, suggests problems with the model. 
For the data and posterior distribution shown in Figure \ref{fig:constraints}, the
Bayesian $p$-value from the PPC for the HI mass function is $p_{\rm HI}=0.25$, 
which indicates an adequate fit, but for the $K$-band luminosity function it
is only $p_{\rm K}=0.011$. This is much lower than
the value of $p_{\rm K}=0.66$ obtained in \citet{Lu2012}, which used only
the $K$-band luminosity function as the data constraint.
Hence, the model constrained using both the HI mass function
and the $K$-band luminosity function \emph{does not} fit the
observational data well.  In the remainder of this paper we will 
explore the causes for this discrepancy and their implication 
for models of galaxy formation generally.

\begin{figure*}
 \centering
 \includegraphics[width=1.0\textwidth]{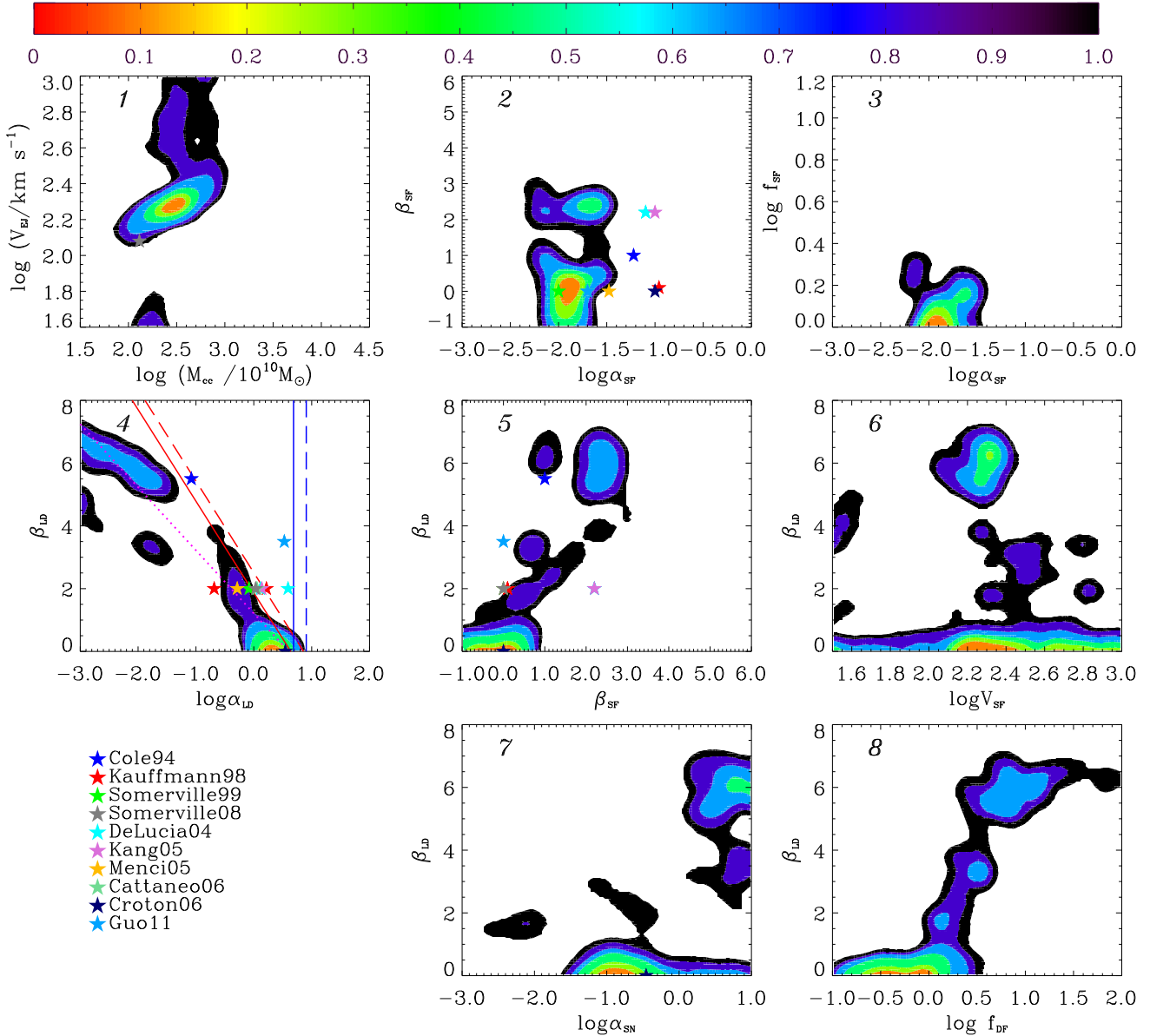}
 \caption{The marginalised posterior distributions of 10 model
   parameters. Panel 1: AGN feedback shuts off gas cooling for haloes
   with $M_{\rm vir}>M_{\rm CC}$ and winds expel gas from haloes with
   circular velocities $\vvir>V_{\rm EJ}$. Panel 2: $\alpha_{\rm SF}$ 
   parametrises the normalisation of star formation efficiency, 
   which is a constant for haloes with a circular velocity higher 
   than parameter $V_{\rm SF}$, and a power-law 
   function of $\vvir$ with $\beta_{\rm SF}$ being the power law
   index of the relation. Panel 3: $f_{\rm SF}$ is the cold
   gas surface density threshold for star formation and $\alpha_{\rm SF}$ is
   as above. Panel 4: $\alpha_{\rm
     LD}$ parametrises the loading factor of the SN feedback and
   $\beta_{\rm LD}$ is the power index of the power-law dependence of
   the loading factor on the halo circular velocity.  The blue lines
   denote the kinetic energy limit based on Type II SN for
   outflows in haloes with a circular velocity of 220 km/s for two IMFs:
   Salpeter (solid) and Chabrier (dashed).  The red lines show the
   extreme parameter combination of the two mass-loading parameters
   for haloes with a circular velocity of 100 km/s. Models in regions to
   the right of the limiting lines require more than
   100\% of the total available energy for the cold gas ejection.  
   The magenta dotted line denotes models with a loading factor 
   equal to 4 for haloes with a circular velocity of 70 km/s.  
   Panel 5: the joint marginalised posterior distribution in 
   the $\beta_{\rm LD}$-$\beta_{\rm SF}$ plane.      
   Panel 6: the joint marginalised posterior distribution 
   in the $\beta_{\rm LD}$-$V_{\rm SF}$ plane.    
   Panel 7: $\alpha_{\rm SN}$ characterises the fraction of SN energy that
   is required to power the ejection of hot halo gas.  
   Panel 8: the joint marginalised posterior distribution in the 
   $\beta_{\rm LD}$-$\log f{\rm DF}$ plane, 
   where $f_{\rm DF}$ is a coefficient characterising the 
   satellite dynamical friction timescale. 
   The greyscale (colour in on-line version) denotes the confidence levels as 
   shown by the colour bar at the top of the figure. 
   Symbols with different colours in the various panels denote
   the corresponding parameters adopted in some earlier SAMs, as indicated. 
   They do not appear in every panel for all models because 
   some parameters are absent in the parametrisation of those models. 
  }
 \label{fig:posterior}
\end{figure*}

Figure \ref{fig:posterior} shows marginalised posterior
distributions for 10 parameters characterising star formation
and feedback and we now describe those with marginal distributions that
differ significantly from those in \citet{Lu2012}.  In the upper-left panel
of Figure \ref{fig:posterior},  the vertical axis shows the model parameter
$V_{\rm EJ}$: the halo circular velocity below which all outflowing
mass reheated from the disk
must be ejected out of the host halo. The horizontal axis
shows the parameter $M_{\rm CC}$: the critical halo mass above which
radiative cooling of the halo gas is shut off to mimic
radio-mode AGN feedback \citep{Croton2006,Bower2006}.
The mass $M_{\rm CC}$ is well constrained, with a value about $10^{12.5}\msun$.
This result also agrees with the value obtained by \citet{Lu2012} using
the $K$-band luminosity function alone as a data constraint, and with
other existing SAMs 
that implement ``halo quenching'' \citep[e.g.][]{Cattaneo2006, Somerville2008}. 
Including the HI mass function as an additional
constraint does not affect the inferred value of $M_{\rm CC}$ significantly
because $M_{\rm CC}$ controls gas cooling and star formation in
massive haloes while the HI mass function mainly constrains low-mass
haloes, which contain most of the HI gas.

The value of $V_{\rm EJ}$ is also well constrained, with $\log V_{\rm
  EJ}\approx 2.3$ ($V_{\rm EJ}\approx 200\,{\rm km/s}$), suggesting
that all haloes with circular velocities lower than this are required to
have efficient mass ejection to fit the data, and the same ejection process 
can be inefficient in haloes with higher circular velocities. There is also a non-negligible  
tail to even larger $V_{\rm EJ}$; models in this tail require efficient mass 
ejection at even higher halo circular velocities.  
This result agrees with both the model of \citet{Somerville2008} and a
recent analysis of the Munich model \citep{Henriques2013}: efficient gas
ejection from halos is required to match multiple data sets simultaneously.  
Without the HI constraint, the
observed $K$-band luminosity function of galaxies can be reproduced by
making the star formation efficiency a strong function of
circular velocity, resulting in a massive cold-gas component.
However, this makes the amplitude of the corresponding predicted HI mass
function an order of magnitude higher than that observed
\citep{Lu2012, Wang2012}.  In our model, the strong degeneracy between outflow
and star formation efficiency is broken by the constraint provided by
the HI mass function.  Consequently, one requires efficient gas ejection from
halos to prevent a large reservoir of cold gas and only requires
the star formation efficiency to be a weak function of
halo circular velocity (see below). If we fix $V_{\rm EJ}=0$, we
find that the model can no longer match the observational data;  
the best fitting model, i.e. in this case
the model with the maximum a posteriori (MAP) 
probability, has a likelihood that is almost 9,000,000 times lower than our
fiducial MAP.  
This suggests that efficient gas ejection is unavoidable in our model family.
 
\subsubsection{Star formation}

With the HI mass function as an additional constraint, the posterior
probability of the star formation parameters is now constrained to a
fairly narrow range (see Panel 2 of Figure~\ref{fig:posterior}).
The posterior of $\log(\alpha_{\rm SF})$ peaks sharply at $-1.8$, implying
that a galaxy converts about $\approx 1.6\%$ of its cold gas into
stars in every dynamical time.  This is consistent with the observed efficiency
in galaxies \citep{Silk1997, Elmegreen1997}.  The posterior
for $\beta_{\rm SF}$ is now peaked near zero, much lower than the
value of $\beta_{\rm SF}\approx10$ obtained in \citet{Lu2012} constrained using
the $K$-band luminosity function alone. This low value for $\beta_{\rm
  SF}$ means that the star-formation rate is largely determined by the
available supply of cold dense gas and does not depend on the mass (or
circular velocity) of the host halo.  This is consistent with
observed star-formation laws \citep{Schmidt1959, Kennicutt1989}, in which the
star formation rate depends only on gas properties. However, there still remains
a weaker mode - its posterior odds relative to the main mode is 2.5 times
smaller - where $\alpha_{\rm SF} \approx 2$\%  and $\beta_{\rm SF}\approx 2.5$.
We will show that the parameter $\beta_{\rm SF}$ is covariant with 
feedback parameters in another panel of the figure. 

In \citet{Lu2012}, the star-formation efficiency parameter
$\alpha_{\rm SF}$ was found to be strongly covariant with the
threshold of cold gas surface density for star formation, $f_{\rm SF}$:
a high star formation efficiency with a high threshold
value was equivalent to a low star formation efficiency with a low
threshold value.  As expected, the addition of the HI mass function breaks
this degeneracy.  Panel 3 of Figure \ref{fig:posterior} shows the
marginalised posterior probability distribution in the $\log f_{\rm
 SF}$--$\log \alpha_{\rm SF}$ plane.  Remember that in our parametrisation,
 $f_{\rm SF}$ is equal to the surface gas density threshold for star
formation, $\Sigma_{\rm SF}$, under the assumption that $r_{\rm
  disc}=r_{\rm disc,0}$, i.e.  that the disc size is given by the model of
\citet{Mo1998}.  The combined data sets require that the threshold surface
density for star formation is small; we find
$\Sigma_{\rm SF}\approx 1 \msun {\rm pc}^{-2}$. 
This value is about one order of magnitude lower than the
observed threshold surface density \citep{Kennicutt1998, Bigiel2008, Leroy2008}
and is also right against the prior bound, so even lower values may be preferred
by the model, although such low values would be in even greater conflict with
observations.  In the model, a high gas surface density threshold would
result in too much cold gas in low-mass haloes at large radii and too little
star formation to power enough outflows.  This problem was highlighted by
\citet{Mo2005}, who found that the standard threshold surface density
for star formation would lead to an HI mass function that was too high, even
if each low-mass halo only contained a cold gas disk at the threshold surface
density.

Our SAM requires both powerful, efficient feedback to remove the gas from the
halo and a threshold gas surface density that is much lower than the
observed star formation threshold (and still fails to match the data).
To explore explicitly the sensitivity to a
higher surface density threshold, we constrained the prior of $f_{\rm SF}$ to
range between $3\msun\, {\rm pc}^{-2}$ and $30\msun\, {\rm pc}^{-2}$.
We find that this greatly reduces the goodness of fit and that the
best fits are always obtained with $f_{\rm SF}$ at the lower boundary of
the prior. The MAP using this restricted prior has a
likelihood that is smaller than the MAP in the
fiducial run by a factor of over 3,000,000.  If we further restrict $f_{\rm SF}$
to match the observed value of $10 \msun\,{\rm pc}^{-2}$, the likelihood
decreases by a further factor of over 60,000,000,
and the $p$-value for the HI-mass function becomes very 
small, $p_{\rm HI}<0.005$, indicating an unacceptable fit.  
Thus, the current model family cannot explain the observed HI 
mass function using the observed gas density threshold
for star formation.  

\subsubsection{Feedback}

Panel 4 of Figure \ref{fig:posterior} shows the marginalised
posterior probability distribution of the two supernova gas ejection parameters:
$\alpha_{\rm LD}$ and $\beta_{\rm LD}$. 
For comparison, we also plot the combinations of these parameters used in a 
number of existing SAMs.
These models include \citet{Cole1994}, \citet{Kauffmann1998a},
\citet{Somerville1999}, \citet{Somerville2008}, \citet{DeLucia2004}, \citet{Kang2005},
\citet{Menci2005}, \citet{Cattaneo2006}, and \citet{Guo2011}.  
The parameter choices are converted into our
parametrisation, where the amplitude of the mass-loading factor is
normalised at a halo circular velocity scale of 220 km/s. As expected,
because of the similarity in our phenomenological parametrisations, most
of the other SAMs are located in the high probability region of our
inferred posterior.  These two parameters are covariant and populate three 
main modes: $\beta_{\rm LD} \approx 0$ that is equivalent to the Croton
model \citep{Croton2006}, $\beta_{\rm LD} \approx 2$ that is similar to the
Somerville model \citep{Somerville2008} and a number of other models, and
a large value of $\beta_{\rm LD} \approx 6$ \citep[e.g.][]{Cole1994}.  

Because we parametrise the outflow loading factor as a power-law
function of halo circular velocity, $\alpha_{\rm ld}=\alpha_{\rm
  LD}(\vvir/220{\rm km\,s^{-1}})^{-\beta_{\rm LD}}$,  a covariance
represented by a straight line in the $\log \alpha_{\rm
  LD}$--$\beta_{\rm LD}$ plane corresponds to a group of models that
have a fixed loading factor for a specific circular velocity.  The
magenta dotted line in Panel 4 of Figure
\ref{fig:posterior} roughly captures the covariance of the two
parameters and corresponds to a loading factor of 4 for haloes
with a circular velocity of $70{\rm km/s}$ or a mass of about $1.5\times
10^{11}\msun$ at the present time.  The covariance suggests that 
these haloes may be critical for fitting the $K$-band luminosity function 
and the HI mass function of galaxies simultaneously.

Our fiducial model does not limit the total energy available for feedback
but, of course, Nature does impose limits. 
The total energy available for outflows powered by Type II
supernovae is
\begin{equation}
E_{\rm SN, max}= \phi \eta_{\rm SN} \epsilon_{\rm SN}\,,
\end{equation}
where $\epsilon_{\rm SN}\sim 10^{51}{\rm ergs} =5 \times 10^7\msun
{\rm km^2\,s^{-2}}$ is the total energy output of a SN, and $\eta_{\rm
  SN}$ is the number of SNs expected from the formation of one solar
mass of stars.  Assuming that Type II SN progenitors are stars with
masses $8\msun <m <20\msun$, then $\eta_{\rm
  SN}=5.88\times10^{-3}\msun^{-1}$ for a Salpeter IMF
\citep{Salpeter1955}, and $\eta_{\rm SN}=9.76\times10^{-3}\msun^{-1}$
for a Chabrier IMF \citep{Chabrier2003}.  The total energy consumed by
the outflow depends on the detailed processes and the final state of
the outflow material.  If the outflow is thermalised and settles into
an equilibrium state in the gravitational potential of the host halo,
the energy required is ${5 \over 4} m_{\rm rh} \vvir^2$, where $m_{\rm
  rh}$ is the mass of the reheated gas \citep[e.g.][]{Kauffmann1998a}.
Since only a fraction of the energy is expected to heat the gas:
\begin{equation}
  \eta_{\rm SN} \epsilon_{\rm SN} \geq 
  {5 \over 4} \alpha_{\rm LD} \left({220{\rm km/s} \over
      \vvir}\right)^{\beta_{\rm LD}} \vvir^2.
\end{equation} 
Thus, for haloes with $\vvir = 220\,{\rm km/s}$, the maximum
normalisation of the loading factor, $\alpha_{\rm LD, max}(\vvir=220)
=4.13 \times {\eta_{\rm SN} \over 5\times 10^{-3}}$, which
is 4.8 for a Salpeter IMF and 8.0 for a Chabrier IMF.
These values are plotted in the lower-right panel of Figure
\ref{fig:posterior} as the blue solid line and the blue dashed line,
respectively.

If the outflow is ejected directly from the centre of the host
halo, the energy required to escape is ${1 \over 2} m_{\rm ej} V_{\rm esc}^2$,
where $V_{\rm esc}$ is the escape velocity. For a NFW halo
\citep{Navarro1996} with a concentration $c=10$, the escape velocity from
the centre is $V_{\rm esc}^2\approx 13\times\vvir^2$.  
Thus, to eject the gas from such a halo, the lower
limit on the required energy is $E={13 \over 2} m_{\rm LD} \vvir^2$,
where $m_{\rm LD}$ is the total gas mass loaded in the wind.  The mass-loading
factor in such haloes is, therefore, limited by the total energy
provided by SN explosions, and the solid red line and the red dashed
line in Panel 4 of Figure~\ref{fig:posterior} show these
limits for a Salpeter and a Chabrier IMF, respectively. Models
lying above such a limiting line need more energy than is available from
haloes with circular velocities equal to or below 100 km/s.  The
posterior contours for all the modes with $\beta<4$ are $\sim 0.3-0.6$ dex 
to the left of but not far from the energy limits for haloes with 
$\vvir\leq 100\, {\rm km/s}$, indicating that a large fraction, i.e. 25--50\%,
of all the supernova energy has to be put into the outflow in such galaxies.

The required high loading factor results from the need to simultaneously match
the shallow low-end slope of both the $K$-band luminosity function and the HI
mass function.  The total baryonic mass that can be accreted into a
halo is $f_{\rm b}\mvir$, where $f_{\rm b}$ is the universal baryon
fraction. Suppose a fraction $f_{\rm c}$ of the gas can cool and
collapse onto the central galaxy.  For low mass haloes with $\vvir\la
150$ km/s, $f_{\rm c}$ is expected to be close to unity because of the
high cooling efficiency \citep{Thoul1995, Lu2007}.  If the expelled
gas can fall back onto the galaxy and go through multiple cycles, the
effective $f_{\rm c}$ can be larger than 1.  However, to explain the
shallow low-mass end of the HI mass function and the faint end slope
of the $K$-band luminosity function, the cold gas and stellar mass has
to be very low relative to $f_{\rm b}\mvir$. Indeed, the stellar mass
to halo mass ratio for haloes with masses $\sim 10^{11}\msun$ is only
about $3\times10^{-3}$ \citep{Yang2012,Papastergis2012, Behroozi2013}, 
and the cold gas mass to stellar mass ratio is no
more than 10 \citep{Kannappan2004, Papastergis2012}. Thus, the gas
associated with low-mass haloes either is never accreted into the
galaxy or has been expelled by galactic winds. In the latter case, if we
conservatively assume that the cold gas mass is 10 times the stellar
mass, then the total cold baryonic mass in a $10^{11}\msun$ halo is
$\sim 0.033$ of the halo mass.  If we make the further conservative assumption
that for every unit of stellar mass formed at early times only half of it
remains in stars today owing to stellar mass loss, the total mass that was
ever involved in star formation is $\sim 0.006$ of the final halo mass.
Based on these conservative assumptions, one finds that the mass-loading 
factor is $(f_{\rm b}-0.033)/0.006 = 23$, which is close to the upper limit
obtained from the total kinetic energy expected from SN. 

The posterior marginals (Fig. \ref{fig:posterior}) imply that 25--50\% of the
available SN energy, depending on the assumed IMF, is required to power the
wind.  This required high efficiency of energy loading poses a severe challenge
for the standard SN feedback scenario
\citep[e.g.][]{Dekel1986, White1992}.  Indeed, detailed hydrodynamical
simulations by \citet{MacLow1999} and \citet{Strickland2000}
demonstrated that supernova feedback is very inefficient in expelling
mass because the onset of Rayleigh-Taylor instabilities severely
limits the mass-loading efficiency of galactic winds.  There are
a couple of possible solutions to this dilemma.  First, a top-heavy or 
bottom-light IMF would yield a larger $\eta_{\rm SN}$. Our data constraints
are from the local galaxy population, 
which by themselves do not distinguish different IMFs.  
Second, the UV photons from massive stars could effectively transfer radiation
momenta to the gas via dust opacity, thereby producing momentum driven winds
\citep{Murray2005}.  The total
radiation energy available is about 100 times as high as the kinetic
energy of supernova explosions, vitiating the mechanical energy limit.
The viability of this mechanism is uncertain: while some simulations
show that it is effective \citep[e.g.][]{Hopkins2011b}, others find
that the development of Rayleigh-Taylor type instabilities can suppress
the formation of outflows \citep{Krumholz2012b}. Thus, although the
required high efficiency may rule out mechanisms based on SN
energy-driven winds alone, the viability including momentum-driven winds is
still unclear.

The star formation feedback models implemented in SAMs are crude.  In
our model, the outflow reduces the total cold gas mass of the galaxy
and the remaining cold gas is redistributed in an exponential profile
at each time step. This is similar to some cosmological hydrodynamical
simulations \citep[e.g.][]{Dave2013} where the effective viscosity of the disk
gas, which could include actual gas viscosity and/or gravitational torques,
causes the cold gas from the outer disk to move rapidly inwards to
replenish the star forming gas, and so star formation can continue
near the disk centre even if the total amount of gas in the disk is
reduced.  However, since the gas viscosity in such simulations is
artificial and the resolution is still limited, it is unclear whether the
effect is physical or numerical. For example, if the effective viscosity of the
disk gas were unimportant, much of the cold gas would stay in the outer
region of the disk without forming stars.  Some simulations
of individual disks \citep{Forbes2012} do find that star formation
feedback is localised to regions where the star formation is active,
producing a disk with a reduced cold gas density only in the central
part of the disk, leaving the cold gas in the outskirts intact.  Others confirm
the cosmological simulation results \citep[e.g.][]{Hirschmann2013}.  To
explore this in our SAM, we have considered a model in which star
formation feedback is assumed to only affect the gas within the radius
where star formation can happen, i.e. the cold density is above the
threshold surface density.  Without a lower limit on $f_{\rm SF}$, the
posterior distribution is nearly unchanged owing to the low surface
density threshold for star formation and the high outflow rate
required by the shallow low-mass end of the galaxy luminosity
function. However, if the value of $f_{\rm SF}$ is forced to match the
observed star formation threshold surface density of $10\msun\,
{\rm pc}^{-2}$ then the model fails to fit the data ($p_{\rm HI}<0.005$), 
consistent with the results obtained by \citet{Mo2005}.

\begin{figure}
  \centering
  \includegraphics[width=0.5\textwidth]{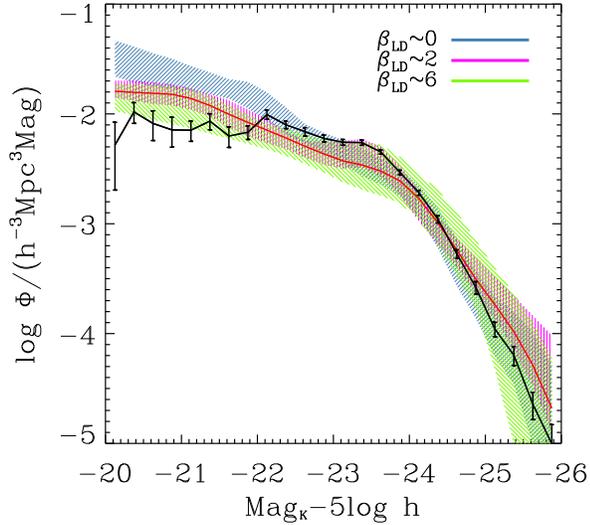}
  \caption{The predicted $K$-band luminosity function for the three different
    modes present in the lower-right panel of Fig.~\ref{fig:posterior}
    represented by different values of $\beta_{\rm LD}$.  The
    red line is the median prediction for the entire posterior.
      }\label{fig:fend}
\end{figure}

As we remarked before, we find that there are three major modes in the
$\log \alpha_{\rm LD}$--$\beta_{\rm LD}$ plane located at $\beta_{\rm
  LD}\approx$ 0, 2, and 6, as one can see in Panel 4 of
Figure~\ref{fig:posterior}.  The relative posterior odds in each mode
are 4:1:2, respectively.  Hence, the $\beta = 0$ mode is favoured by
the combined data sets in terms of marginalised posterior probability.
Since our prior distribution is intentionally very broad, as we
describe in \S\ref{sec:model}, and this will quantitatively affect
these odds.  That said, the maximum likelihood is located in the
$\beta_{\rm LD}\approx6$ mode, and the highest likelihood values in
the $\beta_{\rm LD}\approx2$ and 0 modes are 30 and 1,400 times lower,
indicating that those two modes produce a worse fit to the data than
the $\beta_{\rm LD}\approx6$ mode.  
The PPC $p$-values calculated for the $K$-band luminosity function predicted 
by each of the modes also indicate that the $\beta_{\rm LD}\approx6$ mode 
produces a better, but still inadequate, fit to the data,  
$p_{\rm K}=0.016$ for the $\beta_{\rm LD}\approx6$ mode, and $p_{\rm K}<0.005$ for the other two modes. 
In Figure\,\ref{fig:fend}, we show
the $K$-band luminosity function produced by the three modes. The
shaded blue region shows the posterior predictions of models with
$\beta_{\rm LD}<1.25$, the magenta region shows the posterior
predictions of models with $1.25 < \beta_{\rm LD}<4$, and the green
region shows the posterior predictions of models with $\beta_{\rm LD}>4$.
Note that the three modes produce indistinguishable
predictions for the HI mass function, as the predictive distribution
of the HI mass function produced by the entire posterior shown in
Figure \ref{fig:constraints} is narrow.  The predicted faint-end slope
of the $K$-band luminosity function is steeper for a smaller value of
$\beta_{\rm LD}$, and the current data apparently favours the mode
with the highest $\beta_{\rm LD}$. This trend confirms recent
hydro-dynamical simulation results of \citet{Puchwein2013} and
\citet{Dave2013}.  Moreover, this demonstrates that discrimination
between these modes could be possible given improved data for the
faint end of the luminosity function.

Even though the $\beta_{\rm LD}\approx 0$ mode has relatively lower
likelihood, it has the highest marginal posterior probability.  This implies
that the modes with high values of $\beta_{\rm LD}$ are supported by
proportionately smaller volumes in parameter space relative to the
$\beta_{\rm LD}\approx0$ mode.  This could be investigated in detail
using the distribution of marginal likelihood, but the required
resolution to compute the volume in the high-dimensional parameter space is computationally
prohibitive.  Nonetheless, relying on the posterior marginals shown in
Figure \ref{fig:posterior}, we believe that the complex posterior results from the fine
tuning necessary to simultaneously match the $K$-band luminosity and
HI mass function.  Specifically, Panel 5 indicates that $\beta_{\rm
  SF}$ correlates with $\beta_{\rm LD}$.  As $\beta_{\rm SF}$
increases, the increasingly smaller star formation efficiency in low
circular velocity haloes leaves more cold gas in the disk. To
simultaneously match the HI mass function, the model thus needs to
increase the feedback efficiency in lower circular velocity halos by
increasing $\beta_{\rm LD}$.

In addition, sometimes a particular parametrisation can also introduce 
specific features into the posterior. 
Panel 6 of Fig. \ref{fig:posterior} shows an example; the parameter
$V_{\rm SF}$ is not constrained
when $\beta_{\rm LD}\approx0$, but is strongly constrained when
$\beta_{\rm LD}\approx6$.  When $\beta_{\rm LD}\approx0$, $\beta_{\rm SF}$ is
also about 0, meaning a constant star 
formation efficiency. In this case, as described in \S\ref{sec:model}, 
$V_{\rm SF}$ no longer has any effect on the model predictions, because 
the parametrisation for the star formation efficiency no longer 
has a transition scale that $V_{\rm SF}$ describes.  
Thus, the likelihood is the same no matter what value 
$V_{\rm SF}$ takes when $\beta_{\rm LD}\approx 0$.

Moreover, the high $\beta_{\rm LD}$ mode requires two unusual
conditions. First, as shown in Panel 7 of Fig. \ref{fig:posterior},
for $\beta_{\rm LD}\approx 6$, the parameter $\alpha_{\rm SN}$, which
characterises the fraction of SN kinetic energy used to eject the hot
halo gas, is always required to be higher than one, implying that the
required energy is higher than the kinetic energy provided by SN Type
II. The reason is that $\beta_{\rm LD}\approx 6$ leads to a rapid
decrease of cold gas ejection with increasing halo circular velocity,
and a strong outflow of hot halo gas in relatively massive halos is
needed by the mode to fit the data.  This conclusion agrees with the
result of \citet{Mutch2013} that one needs more energy than that
available from SN Type II to explain the evolution of the galaxy
stellar mass function from $z\approx 0.8$ to 0.  Second, the
$\beta_{\rm LD}\approx 6$ mode also requires a long time scale, about
10 times as long as the fiducial dynamical friction timescale, for
satellite galaxies to orbit in the host halo before merging into
central galaxies (see Panel 8 of Fig. \ref{fig:posterior}).  This can
be understood as follows. When $\beta_{\rm LD}$ is large, the effect
of star formation feedback is increasingly weaker for higher mass
haloes. In this case, the merging timescale for satellite galaxies is
required to be long to prevent them from merging into the halo centre
and hence significantly increasing the mass of the central
galaxy. Since a longer merger time also leads to a higher satellite
fraction, observational data of satellite fraction can tighten the
constraint.  We will come back to this topic in a future paper.

\subsection{Posterior Predictions}
\label{sec:predictions}

The model family generally requires efficient outflows from low-mass
galaxies for its best fit, but even that is not enough to match both of the
constraining observational data sets.  We now use the posterior
distribution to predict the outflow rate and other observables to
investigate the failure of the model family.  Since the marginalised posterior in the
$\alpha_{\rm LD}$--$\beta_{\rm LD}$ plane is multi-modal, we
distinguish the modes according to their value of $\beta_{\rm LD}$ when
making model predictions.  We marginalise the posterior probability to
generate predictions as described in \citet{Lu2012}.

\begin{figure}
  \centering
  \includegraphics[width=0.5\textwidth]{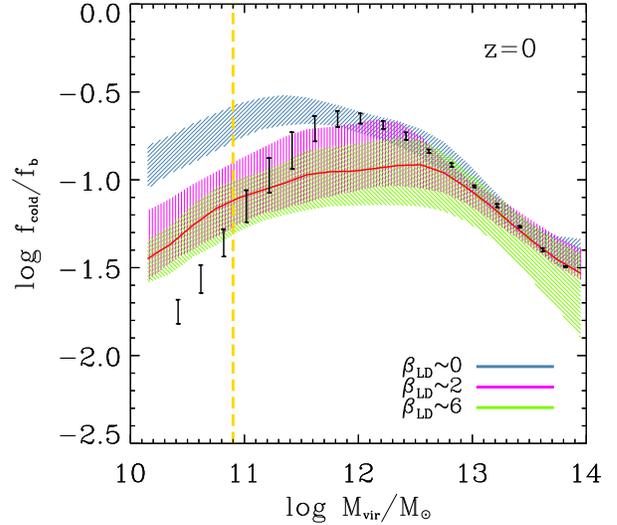}
  \caption{The posterior predictive distribution of the cold baryon
    (stars plus cold gas) mass fraction as a function of halo
    mass. The shaded bands with different colours encompass the 95\%
    credible range of the prediction for the three modes represented by
    different values of $\beta_{\rm LD}$, while the red line is the
    predicted median of the entire posterior predictive distribution.
    The black points with error bars are from \citet{Papastergis2012}.
    The vertical yellow dashed line marks the lowest halo mass
    constrained by the $K$-band luminosity function.
    }
  \label{fig:bfrac}
\end{figure}

Figure \ref{fig:bfrac} shows the posterior prediction for the cold
baryonic mass fraction (cold gas plus stars) of central galaxies
normalised by the cosmic baryon fraction, $f_{\rm b}$, as a function
of halo mass. The red line shows the median of the whole posterior
prediction, while the shaded regions with different colours encompass the
95\% credible ranges from the modes with $\beta_{\rm
  LD}\sim0$, 2, and 6, as indicated in the figure. The
black bars show the results of abundance matching obtained by
\citet{Papastergis2012} using the ALFALFA and SDSS data.

The observed $K$-band luminosity function only determines absolute magnitudes up
to $K={\rm Mag_K}-5\log h=-20$, which only constrains the halo virial mass
above $\sim10^{11}\msun$
(indicated as the vertical line in Fig.\,\ref{fig:bfrac}).  The 95\%
range of the posterior prediction contains the abundance matching
result for haloes more massive than a few times $10^{11}\msun$, but all
three modes overpredict the cold baryon fraction for lower mass
haloes.  Among the three
modes, models with larger $\beta_{\rm LD}$ produce lower cold baryon
mass fractions for low mass haloes. The mode with $\beta_{\rm LD}\approx
0$ vastly overpredicts the cold baryon mass fraction in haloes with
masses $\leq10^{11.5}\msun$.

\begin{figure}
  \centering
  \includegraphics[width=0.5\textwidth]{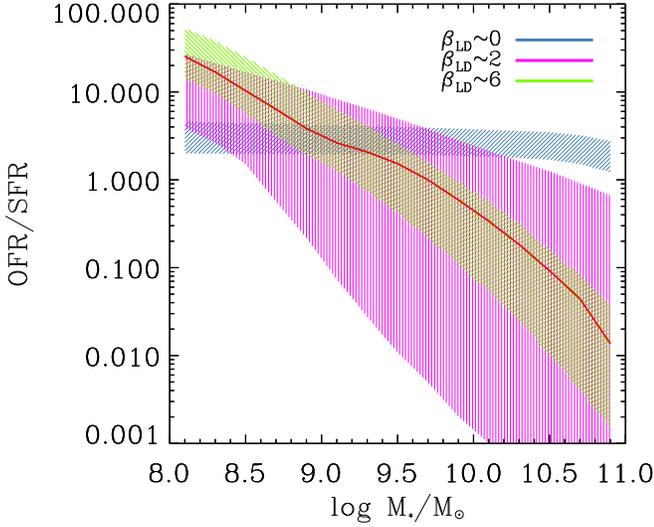}
  \caption{The predicted ratio of mass outflow rate to star formation rate as a
    function of stellar mass for central galaxies at $z=0$.  The
    colour bands and lines are as described in
    Fig. \protect{\ref{fig:bfrac}}.
    }
\label{fig:ofr}
\end{figure}

To explore the physical implications of feedback, we predicted the
average star formation rate (SFR) and outflow rate (OFR) in each mass
bin for all central galaxies at $z=0$.  Figure \ref{fig:ofr} shows the
ratio OFR/SFR, which is commonly referred to as the mass-loading
factor of the wind. Again the red lines show the medians of the whole posterior
prediction, while the shaded regions with different colours encompass
the 95\% credible range from the three modes defined
by their values of $\beta_{\rm LD}$.  The predicted OFR/SFR ratio
generally decreases with stellar mass; it is approximately 10 for
galaxies with stellar masses $\la 10^{8.5}\msun$ and close to 1 for
Milky-Way mass galaxies.  Observationally, there is no strong evidence
for the existence of such a strong stellar mass dependence for 
outflows in normal disk galaxies at the
present day \citep[see][]{Heckman2003, Veilleux2005, Rupke2002,
  Martin2006}.  In starburst galaxies, the mass outflow rates are
roughly comparable to the star-formation rate, so that OFR/SFR $\sim
1$, and lower mass galaxies do not show stronger outflows.
Thus, the high OFR/SFR ratio predicted for the general dwarf
galaxy population does not have any observational support, but then again
there is no strong observational evidence against it.  Comparing the
predictions for the different posterior modes, we see that different
choices of $\beta_{\rm LD}$ produce very different trends of the
loading factor with galaxy stellar mass.  The $\beta_{\rm LD}\approx
0$ mode produces a constant mass-loading factor, ${\rm OFR/SFR}\approx
3$ over a large stellar mass range, while the other two modes predict
a rapidly decreasing ${\rm OFR/SFR}$ with increasing stellar mass.  The roughly
constant ${\rm OFR/SFR}$ predicted by the $\beta_{\rm LD}\sim 0$ mode
is consistent with the observational results of \citet{Martin2006},
but this mode significantly overpredicts the cold baryon mass
fraction for low mass haloes.  Clearly, as one needs a large mass-loading 
factor to achieve better fits to the constraining data,
observations of gas outflows in
low-mass galaxies and how they scale with galaxy mass can provide important 
constraints on the feedback model family considered here.

\begin{figure}
  \centering
  \includegraphics[width=0.5\textwidth]{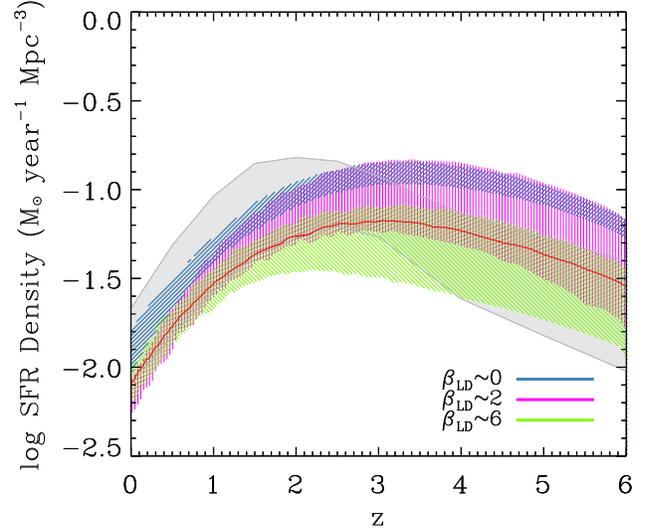}
  \caption{The posterior prediction of the co-moving star formation
    rate density versus redshift.  Points with error bars are
    observational data.  The grey band denotes
    observational estimates compiled in Behroozi et al. 2013. 
    The other colour bands and lines are as described in
    Fig. \protect{\ref{fig:bfrac}}.
    }
  \label{fig:sfrd}
\end{figure}

Figure \ref{fig:sfrd} shows the predicted cosmic star formation rate
density as a function of redshift using the posterior samples selected
from the three modes of $\beta_{\rm LD}$.  
The grey shaded region in the figure denotes a compilation of observational estimates 
of the cosmic star formation rate density as a function of redshift from \citet{Behroozi2013}. 
Compared to the data, our model family predicts a different shape.
The cosmic star formation rate density in the model reaches its 
maximum at higher redshifts ($z\approx3.5$), 
while data suggests it more sharply peaks around $z\approx2$, although the
observed drop at high redshifts could partly owe to missing low mass galaxies
\citep{Lu2014}.
The model also underpredicts the star formation
rate density around $z=2$, as the median of the posterior predictive
distribution is only marginally enclosed by the lower bound of the
observational data. The predictions from the
$\beta_{\rm LD}\approx0$ mode are systematically higher and might be more
consistent with the observations.  The high $\beta_{\rm LD}$ mode
predicts a star formation rate density that is too low to match the
observational data at $z>1$. This stems from the powerful SN driven
outflows in low-mass haloes required to fit the local galaxy luminosity
function and HI mass function. Because the parametrisation for feedback does not 
have an explicit redshift dependence, such strong outflows work throughout all cosmic time in the model. 
This feedback suppresses star formation at high redshift where such low-mass haloes
dominate.

\begin{figure}
  \centering
  \includegraphics[width=0.5\textwidth]{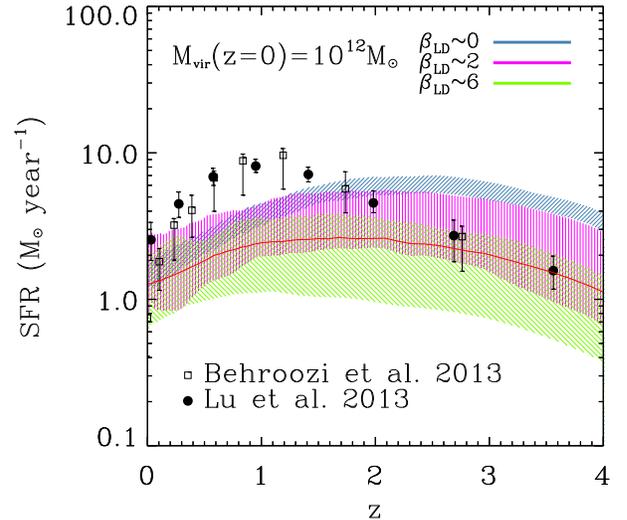}
  \caption{The posterior prediction of the star-formation-rate history
    for haloes with present-day mass $10^{12}{\rm M}_\odot$.
    Points with error bars show results obtained with two recent
    empirical models.  The colour bands and lines are as
    described in Fig. \protect{\ref{fig:bfrac}}.
    }
  \label{fig:sfrd_mw}
\end{figure}

Figure \ref{fig:sfrd_mw} shows the
star formation rate (SFR) as a function of redshift for haloes with
masses of $M_{\rm vir}=10^{12}\msun$ today ($z=0$) for the
three modes in $\beta_{\rm LD}$. The model predictions are compared
with the recent results obtained by \citet{Lu2014} and 
\citet{Behroozi2013} 
using empirical models tuned to fit the observed stellar mass
functions at different redshifts.  The empirical models predict a peak
near $z\approx1$ while our SAM family predicts a much weaker peak at
higher $z$.  This is consistent with the underprediction of the SFR
density at $z\approx1$ in Figure \ref{fig:sfrd}.  The $\beta_{\rm
  LD}=0$ mode also overpredicts the SFR at $z>2$, although this mode
better matches the observed cold gas fraction in haloes with masses of
$10^{12}{\rm M}_\odot$ (see Figure \ref{fig:bfrac}).  Overall,
the predicted star formation histories for Milky Way sized galaxies by
our SAM family are inconsistent with the results obtained from the
empirical models.

\begin{figure}
  \centering
  \includegraphics[width=0.5\textwidth]{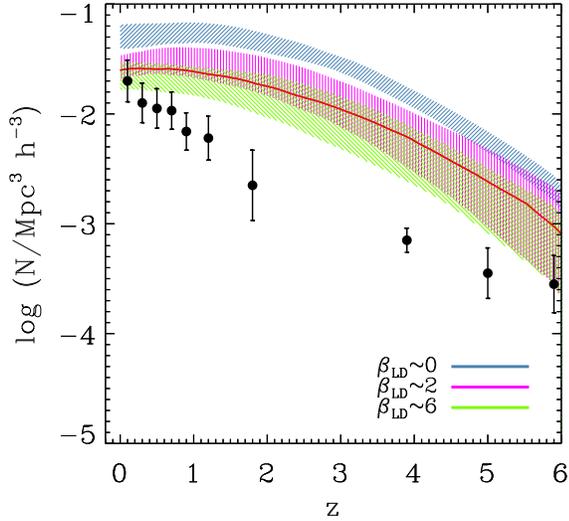}
  \caption{The posterior prediction for the evolution of the number
    density of galaxies with stellar masses in the range
    $\log(M_*/\msun)=9.27-9.77$ as a function of redshift.  Points
    with error bars show data compiled by \citet{Weinmann2012a}.  The
    colour bands and lines are as described in
    Fig. \protect{\ref{fig:bfrac}}.
    }
  \label{fig:ndsm}
\end{figure}

\citet{Weinmann2012a} found that SAMs and numerical
simulations cannot generally reproduce the observed number density evolution of
low mass galaxies.  For comparison, we predict the number densities of
galaxies with stellar masses in the range $9.27\leq\log
M_*/\msun\leq9.77$ from $z=0$ to 6, and compare them with the observational data
compiled by \citet{Weinmann2012a}.  Figure \ref{fig:ndsm} shows that
the model family can accommodate the observational data at $z=0$, but
vastly overpredicts the number density of low-mass galaxies at higher
redshift.  This confirms the results of \citet{Weinmann2012a}, and
suggests that the current model family cannot match the data even if a
large parameter space is explored.

\begin{figure}
\centering
\includegraphics[width=0.5\textwidth]{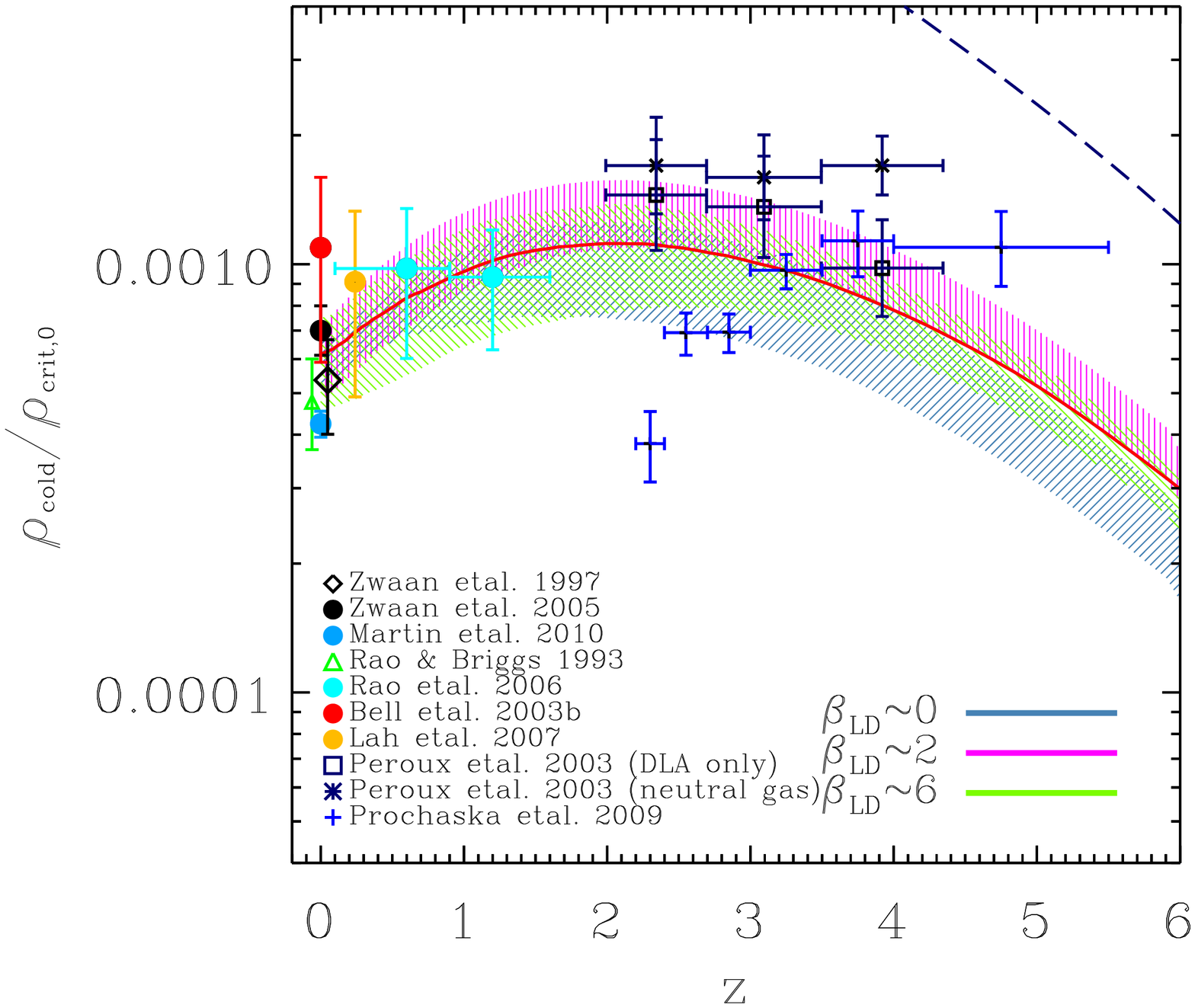}
\caption{The posterior prediction for the co-moving cold gas mass
  density normalised by the critical density of the universe at the
  present time.  The blue dashed line shows the total mass in haloes
  with $\mvir\geq5\times10^9\msunh$ times the universal baryon
  fraction $f_{\rm b}$.  Points with error bars are observational
  data from \citet{Zwaan1997, Zwaan2005, Martin2010, Rao1993, 
  Rao2006, Lah2007, Peroux2003} and \citet{Prochaska2009}.  
  The colour bands and lines are as described in
  Fig. \protect{\ref{fig:bfrac}}.
  }
\label{fig:cgmd}
\end{figure}

Figure \ref{fig:cgmd} shows the prediction of the co-moving mass
density of cold gas as a function of redshift.  
The observational data at high redshifts, estimated using damped Lyman alpha
systems \citep{Peroux2003, Prochaska2009}, suggests
that the cold gas density at high redshifts is higher than in the local
universe.  Our predicted increase of the cold gas mass density with
redshift in the range from $z=0$ to 2 is consistent with the
observations. The model also predicts a decrease of the cold gas
density with increasing redshift at $z>3$. Unfortunately the current
data is still too uncertain to provide meaningful constraints on the
model.

In summary, the model family considered here, which has physical
prescriptions similar to most published SAMs, in addition to not
fitting one of the constraining data sets, the observed $K$-band luminosity
function, also fails to predict additional observations as follows:
\begin{enumerate}
\item It requires very efficient feedback to eject gas from low-mass
  haloes. The energy required is probably too high if the outflows are driven by
  the kinetic energy of supernova explosions alone, although this problem
  may be alleviated by including radiation pressure.
\item It requires a density threshold for star formation that is much
  lower than current observations suggest, owing to the 
  energy needed to eject cold gas from the disk.
\item The high mass-loading factor required by the present-day
  distribution of low-mass galaxies predicts mass outflow rates that
  might be too high to match observation.
\item The predicted star formation rate history of Milky-Way-sized
  galaxies is inconsistent with that obtained from empirical
  models.
\item It overpredicts the number density of low-mass galaxies at
  high redshift.
\end{enumerate}
These results together suggest that some important physics may still be
missing in the current SAMs.

\section{Discussion and conclusions}
\label{sec:discussion}

We use the observed low redshift $K$-band luminosity function and HI-gas mass
function of galaxies as data constraints to explore star-formation and 
feedback in a semi-analytic model
of galaxy formation, which uses MCMC-based Bayesian methodology.  The
combination of these two data sets breaks the degeneracy in the
parametrised phenomenology based on using the $K$-band luminosity function
data constraint alone \citep{Lu2012}.  Our model assumes
that baryons accrete into the assembling halo where they cool,
form stars, and the resulting feedback ejects some fraction of the gas.
Our model has many features in common with most published SAMs and our
results focus on conclusions generic to a large class of published
models. We marginalise over all the uncertainties in the model family that are 
not constrained by the data sets.

Our broad parameter space covers and extends the ranges of the
important parameters for star formation and feedback adopted in many
existing SAMs. Our exhaustive parameter-space search results in the
robust identification of several compact modes that best match the
constraining data.  These modes are likelihood dominated, i.e. the
existence of these modes does not depend sensitively on the prior
distribution as long as the modes are supported.  The analyses
presented in this paper focus on the behaviour of the model
constrained to those modes, namely on the physical implications of the
constrained parameters and the predictions of these modes. These
conclusions are not expected to depend sensitively on the prior.
However, the choice of the prior distribution is generally important in
complex high dimensional problems, especially when the likelihood is
not sufficiently informative.  In such cases, strong prior belief
easily sways weak observational constraints.  Conversely, a thorough
investigation of the poorly constrained model motivates the acquisition
of data that robustly discriminates between competing hypotheses of
interest.

Overall, our model fails to reproduce the joint data sets and,
therefore, more work will be required to understand the nature of the
failure and to propose a remedy before attempting a quantitative Bayesian
analysis.  To assess the fit, we use the posterior predictive check
(PPC) method with a residual sum of squares for each bin, for both the
$K$-band luminosity function and the HI-gas mass function, as a
discrepancy statistic.  In this method, the discrepancy with the data
is compared with the distribution predicted from the posterior
distribution. We find that the data is in the tail of the distribution
defined by the posterior, suggesting the implausibility of the data
given the model.

The primary cause for the failure is that to simultaneously produce the
observed stellar component and the observed cold gas fraction at the low mass
end requires the SN feedback to be
extremely efficient in expelling gas.
This requires a high mass-loading 
factor that consumes much of the available energy budget.  
The low $p$-value we found by exhaustively exploring the parameter space 
has its roots in the over simplicity of the phenomenological model and
suggests that key physical processes are missing or misrepresented.

In addition, we find that the mass-loading factor must depend strongly on
halo circular velocity to obtain the shallow faint-end slope of the luminosity
function.  The model requires that winds from haloes with circular velocities
lower than $\sim 200 {\rm km\,s^{-1}}$ are {\it ejected}  from the halo,
completely.  However, in many mass-loading models the mass-loading 
factor is a weak function of the halo circular velocity.
Recent simulations concur: \citet{Dave2013} show that a steeper
relation between mass-loading and halo circular velocity tends to
produce a shallower stellar and HI mass function at the low-mass end. This
also suggests that accurate observational data for the faint-end slope
of the galaxy luminosity function are crucial to constrain feedback
models.

The inferred requirement for efficient feedback is consistent with the
recent results of \citet{Mutch2013} and \citet{Henriques2013} who also
found that a high efficiency of SN feedback is needed to fit the stellar
mass and/or luminosity functions of galaxies at multiple redshifts.
However, the high efficiency implied by the posterior distribution
obtained here is not supported by hydro-dynamical simulations
\citep[e.g.][]{MacLow1999}, which showed that the feedback efficiency
of SN kinetic energy is usually quite low. Thus, either some other
important energy sources are missing or the way feedback works is
not correctly modelled in the current model family.  For example,
radiation pressure associated with star formation and SN explosions
may provide a viable solution to the energy problem found here
\citep[e.g.][]{Stinson2013}.

Finally, the model predicts the star formation history over all time,
although it is only constrained by data at $z=0$.  The observed star
formation histories of Milky Way-sized haloes and the observed redshift
evolution in the number density of low-mass galaxies at high $z$ are
inconsistent with our posterior predictions.  This suggests that star
formation and feedback may have a more complex redshift dependence
than assumed here.

In summary, our analysis shows that we require a high wind mass loading factor
and mass ejection rate of galactic winds to match the
observed luminosity and cold gas mass functions. This result applies
to many if not all ejective feedback scenarios that depend on winds, largely
independent of the mechanistic details. This is a direct consequence of
a short radiative cooling time scale in low-mass haloes combined with
the small fraction of baryons in stars and cold gas relative to the
cosmic baryon fraction.  Moreover, the predicted high mass-loading factor
and mass-loss rates are not observed, suggesting that
a feedback-generated wind may not be the agent suppressing star formation
and cold gas accretion in low-mass haloes, at least at low-$z$.

Alternatively, this suggests exploring mechanisms that prevent the gas
accretion in the first place.  For example, the intergalactic medium
(IGM) may be preheated by some processes so that low mass haloes can
accrete baryonic matter only at a reduced rate, either by properly
including the effects of already considered processes like supernova winds
or by additional processes like gravitational pancaking or Blazar heating
\citep[e.g.][]{Mo2002, Mo2005, Lu2007, Anderson2010, Zhu2011, Pfrommer2012}.
This may also introduce a characteristic time before and after which the star
formation and feedback proceed differently, as seems to be required to
match the star formation history and number density evolution of low
mass galaxies \citep[see][]{Lu2014}.  We will investigate the effects
of preheating on galaxy formation in a future paper.


\section*{Acknowledgement}

This work is partly supported by grants NSF AST-1109354 (HJM and MDW) 
and NSF AST-0908334 (HJM and NK) and NASA ATP NNX10AJ95G (NK). 
Part of this work used the Extreme Science and Engineering Discovery Environment (XSEDE), 
which is supported by National Science Foundation grant number ACI-1053575.
YL thanks Tom Abel, Peter Behroozi, Eric Bell, Andrew Benson, Darren Croton,
Romeel Dav{\'e},  Andrey Kravtsov,  Emmanouil Papastergis,  Joel Primack, Rachel Somerville,  
Frank van den Bosch and Risa Wechsler for useful discussion. 


\bibliography{bs_h1mf}

\end{document}